\documentclass[11pt, letterpaper]{article}
\usepackage[utf8]{inputenc}
\usepackage{amsmath,amssymb}
\usepackage{graphicx}
\usepackage{subfig}
\usepackage{here}
\usepackage{float}
\usepackage{multirow}
\usepackage[margin=0.75in]{geometry}
\usepackage{color}
\usepackage{ulem}
\usepackage{authblk}
\usepackage{siunitx}
\usepackage{booktabs}
\usepackage{caption}
\DeclareMathAlphabet{\mathpzc}{OT1}{pzc}{m}{it}
\newcommand{\vev}[1]{\langle\Omega|#1|\Omega\rangle}

\renewcommand{\Im}{\mathrm{Im}}

\newcommand{\double}[2]{(#1,\,#2)}
\newcommand{\dif}[1]{d #1}
\newcommand{\glueFourD}{\big\langle \alpha G^{2} \big\rangle}
\newcommand{\glueSixD}{\big\langle g^{3} G^{3} \big\rangle}

\newcommand{\rom}[1]{\uppercase\expandafter{\romannumeral #1\relax}}
\newcommand{\ie}{\textit{i.e.,}}
\newcommand{\eg}{\textit{e.g.,}}

\title{Axial Vector $cc$ and $bb$ Diquark Masses from QCD Laplace Sum-Rules}
\author[3]{S.\ Esau}
\author[1]{A.\ Palameta}
\author[4]{R.T.\ Kleiv}
\author[2]{D.\ Harnett}
\author[1]{T.G.\ Steele}
\affil[1]{Department of Physics and Engineering Physics\\ University of Saskatchewan\\ Saskatoon, SK, S7N 5E2, Canada}
\affil[2]{Department of Physics\\ University of the Fraser Valley\\ Abbotsford, BC, V2S 7M8, Canada}
\affil[3]{Department of Physics\\ University of Waterloo\\ Waterloo, ON, N2L 3G1, Canada}
\affil[4]{Department of Physical Sciences (Physics)\\ Thompson Rivers University\\ Kamloops, BC, V2C~0C8, Canada}
\begin{document}
\maketitle
\begin{abstract}

\noindent Constituent mass predictions for axial vector (\ie\ $J^P=1^+$) 
$cc$ and $bb$ colour-antitriplet diquarks are generated using QCD Laplace sum rules. 
We calculate the diquark correlator within the operator product expansion to next-to-leading-order, 
including terms proportional to the four- and six-dimensional gluon and six-dimensional quark condensates. 
The sum-rules analyses stabilize, and we find that the constituent mass of the $cc$ diquark is $(3.51\pm0.35)$~GeV 
and the constituent mass of the $bb$ diquark is $(8.67\pm0.69)$~GeV. 
Using these diquark constituent masses as inputs, we calculate several tetraquark masses 
within the Type-II diquark-antidiquark tetraquark model.
\end{abstract}

\section{Introduction}\label{I}

Outside-the-quark-model hadrons consisting of four (or more) valence quarks have been theorized for decades. For example, the concept of tetraquarks, hadrons composed of four quarks $(q  q  \overline{q}  \overline{q})$, was introduced in~\cite{PhysRevD.15.267, PhysRevD.15.281} in 1977. Jump forward to 2003 and the discovery of the X(3872) by the Belle collaboration~\cite{Choi:2003ue} and its subsequent confirmation by several other experimental collaborations~\cite{Acosta:2003zx, Abazov:2004kp, Aubert:2004fc, Aaij:2011sn} places us in a new era of hadron spectroscopy. Since then more and more of these hadrons have been discovered in the heavy quarkonium spectra. These hadrons, now collectively referred to as the XYZ resonances, are difficult to explain within the quark model~\cite{Godfrey:1985xj}.
These XYZ resonances have served as a strong motivator for research into beyond-the-quark-model hadrons. See~\cite{PhysRevD.86.010001, Ali:2017jda} for a review of experimental findings and~\cite{Chen:2016qju, Liu:2019zoy} for a review of several multiquark systems.

Looking at four-quark states in particular, there are several  
interpretations of what their internal quark structure might resemble. 
One possibility is that there are no particularly strong correlations between any of the quarks.
However, another possible interpretation is that these states could be 
meson-meson molecule states in which 
two colour-singlet mesons form a weakly bound conglomerate state. 
See~\cite{Close:2003sg, Voloshin:2003nt, Swanson:2003tb, Tornqvist:2004qy, AlFiky:2005jd, Thomas:2008ja, Liu:2008tn, Lee:2009hy} for discussions about  the X(3872) in this configuration. 
Yet another possible interpretation is that  four-quark states are diquark-antidiquark states.
Diquarks are strongly correlated, colour antitriplet pairs of quarks within a hadron. 
(As such, their colour configurations are identical to those of antiquarks.)
See~\cite{RevModPhys.65.1199} for applications of diquarks and~\cite{Jaffe:2004ph} for a discussion of possible diquark configurations. 
In a diquark-antidiquark configuration, the  diquark constituents 
are strongly bound together in a four-quark configuration. 
See~\cite{Maiani:2004vq, Ebert:2005nc, Matheus:2006xi, Terasaki:2007uv, Dubnicka:2010kz} for discussions 
 about the X(3872) in the diquark-antidiquark configuration.
Also, see~\cite{Kleiv:2013dta} for additional discussions on the differences between the molecular and tetraquark models in the context of a QCD sum-rules analysis.

QCD sum-rules analyses of diquarks in several channels have been presented in~\cite{Zhang:2006xp, Dosch1989, Jamin:1989hh, Wang:2011ab, Wang:2010sh, Kim:2011ut}. 
Lattice QCD analyses of light diquarks have also been performed \cite{Hess:1998sd,Alexandrou:2006cq,Bi:2015ifa}.
In this paper, we use QCD Laplace sum-rules (LSRs) to calculate the constituent masses of axial vector 
(\ie\ $J^{P}=1^{+}$) $cc$ and $bb$ diquarks.
The axial vector is the only quantum number that can be realized for  
colour antitriplet diquarks of identical flavours in an S-wave configuration.
We use the operator product expansion (OPE)~\cite{Wilson:1969zs} to compute the correlation function between 
a pair of diquark currents~(\ref{ACurrent})--(\ref{ACurrentDag}). 
In this calculation, in addition to leading-order (LO) pertubative contributions, we also include next-to-leading-order (NLO) perturbative contributions and non-perturbative corrections proportional to the four-dimensional (4d) and 6d gluon condensates as well as the 6d quark condensate. 
The results of these calculations are summarized in Table~\ref{results}.
In particular, we find that the constituent mass of the $cc$ diquark is $(3.51\pm0.35)$~GeV 
and the constituent mass of the $bb$ diquark is $(8.67\pm0.69$)~GeV. 
Substituting these diquark constituent masses into the Type-II diquark-antidiquark tetraquark 
model of Ref.~\cite{Maiani:2014aja}, we calculate masses of several
$[cc][\bar{c}\bar{c}]$, $[cc][\bar{b}\bar{b}]$, and $[bb][\bar{b}\bar{b}]$ tetraquarks.

\section{The Correlator}\label{II}

The axial vector, colour antitriplet diquark current is given by~\cite{Dosch1989,Jamin:1989hh}
\begin{equation}
  j_{\mu, \alpha} = \epsilon_{\alpha\beta\gamma}Q_{\beta}^{T} C\gamma_{\mu}Q_{\gamma}\label{ACurrent}
\end{equation}
with adjoint
\begin{equation}
  j_{\mu, \alpha}^{\dag} = - \epsilon_{\alpha\beta\gamma}\overline{Q}_{\beta}\gamma_{\mu}C \overline{Q}_{\gamma}^{T} \label{ACurrentDag}
\end{equation}
where $C$ denotes the charge conjugation operator, $\epsilon_{\alpha\beta\gamma}$ is a Levi-Civita symbol in quark colour space, and $Q$ is a heavy (charm or bottom) quark field.

Using~(\ref{ACurrentDag}), we consider the diquark correlator
%
\begin{equation}
  \Pi(q^2) = \frac{i}{D-1}
  \left(\frac{q_{\mu}q_{\nu}}{q^2}-g_{\mu\nu}\right)\int d^{D}\!x \;e^{iq\cdot x}
  \vev{\tau[\, j_{\mu,\alpha}(x)\; S_{\alpha\omega}(x,0)\; j_{\nu,\omega}^{\dag}(0)]}
  \label{Correlator}
\end{equation}
%
where $D$ is the spacetime dimension.
In~(\ref{Correlator}), $S_{\alpha\omega}(x,0)$ is a path-ordered exponential, or Schwinger string, given by
\begin{equation} \label{SchwingerString}
S_{\alpha\omega}(x,0) = \hat{\mathcal{P}}\; \text{exp} \bigg[ i g_{s} \frac{\lambda^{a}_{\alpha\omega}}{2} \int_{0}^{x} dz^{\mu} A_{\mu}^{a}(z) \bigg]
\end{equation}
where $\hat{\mathcal{P}}$ is the path-ordering operator. 
The Schwinger string allows gauge-invariant information to be extracted from the gauge-dependent current \eqref{ACurrent}~\cite{Dosch1989, Jamin:1989hh}.   
The explicit cancellation of the gauge parameter has been shown for perturbative contributions up
to NLO~\cite{KleivSteeleZhangEtAl2013}, and in Landau gauge the NLO contributions from the Schwinger string are zero~\cite{Dosch1989, Jamin:1989hh}; hence $S_{\alpha\omega}(x,0) \to \delta^{\alpha\omega}$.  
For non-perturbative contributions of QCD condensates, gauge-invariance of the correlator~\eqref{Correlator} implies that fixed-point gauge methods used to obtain OPE coefficients are equivalent to other 
methods~\cite{Bagan:1992tg}. As observed in Refs.~\cite{Dosch1989, Jamin:1989hh}, the Schwinger string will not contribute to the QCD condensate contributions in the fixed-point gauge, and hence $S_{\alpha\omega}(x,0) \to \delta^{\alpha\omega}$.
Thus, using Landau gauge for pertubative contributions and fixed-point gauge methods for QCD condensate contributions, we can simplify~(\ref{Correlator}) by setting $S_{\alpha\omega}(x,0) \to \delta^{\alpha\omega}$
(as in~\cite{Kleiv:2013dta}).
Lattice QCD analyses of constituent light diquark masses are also based on correlation functions of (coloured) diquark operators~\cite{Hess:1998sd,Alexandrou:2006cq,Bi:2015ifa}.
 Instead of the Schwinger string, gauge dependence of the correlation function is addressed in lattice analyses  either through gauge fixing or coupling to a heavy colour source.

We evaluate the correlator~(\ref{Correlator}) within the OPE to NLO in perturbation theory and include non-perturbative corrections proportional to the 4d and 6d gluon condensates and the 6d quark condensate. 
Each non-perturbative correction is the product of a LO perturbatively computed Wilson coefficient and a QCD condensate. 
The 4d and 6d gluon and 6d quark condensates are defined respectively by
\begin{equation} \label{fourDcond}
\big\langle \alpha G^{2} \big\rangle = \alpha_{s} \big\langle \! \colon \! G_{\omega\phi}^{a} G_{\omega\phi}^{a} \! \colon\! \big\rangle
\end{equation}
\begin{equation} \label{sixDcond}
\begin{aligned}
\big\langle g^{3}G^{3}\big\rangle = g_{s}^{3} f^{abc} \big\langle \! \colon \! G^{a}_{\omega\zeta} \; G^{b}_{\zeta\rho} G^{c}_{\rho\omega} \! \colon\! \big\rangle
\end{aligned}
\end{equation}
\begin{equation} \label{sixDQcond}
\big\langle J^2 \big\rangle
= \frac{D}{6} \, \kappa \, g_s^4 \big\langle \overline{q}q \big\rangle^2
\end{equation}
where $\kappa$ in~(\ref{sixDQcond}) quantifies deviation from vacuum 
saturation. As in~\cite{Palameta:2017ols,Palameta:2018yce}, we set $\kappa=2$ for the remainder of this calculation, 
\eg\ see~\cite{narisonbook:2004} and references contained therein. 

\begin{figure}
\centering
\begin{tabular}{ccc}
\includegraphics[width=50mm]{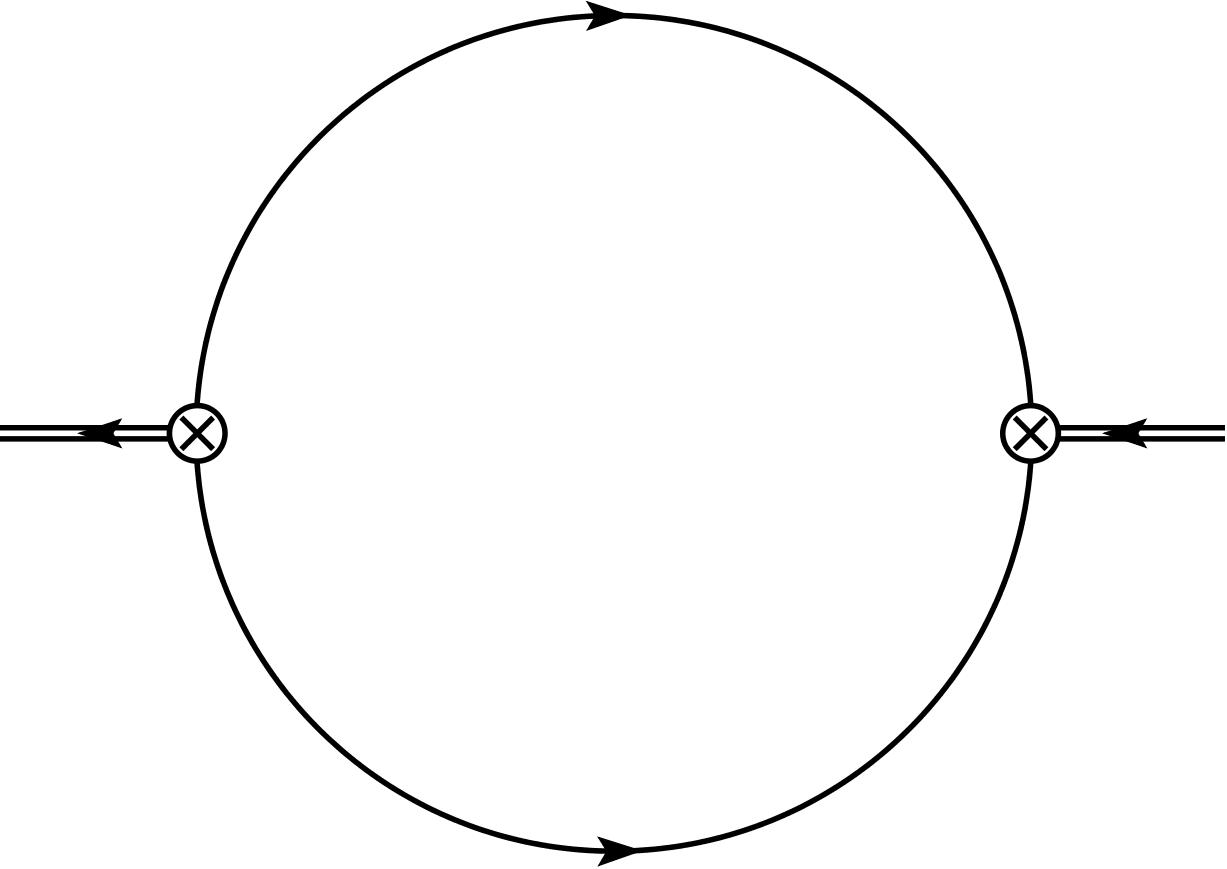} & \includegraphics[width=50mm]{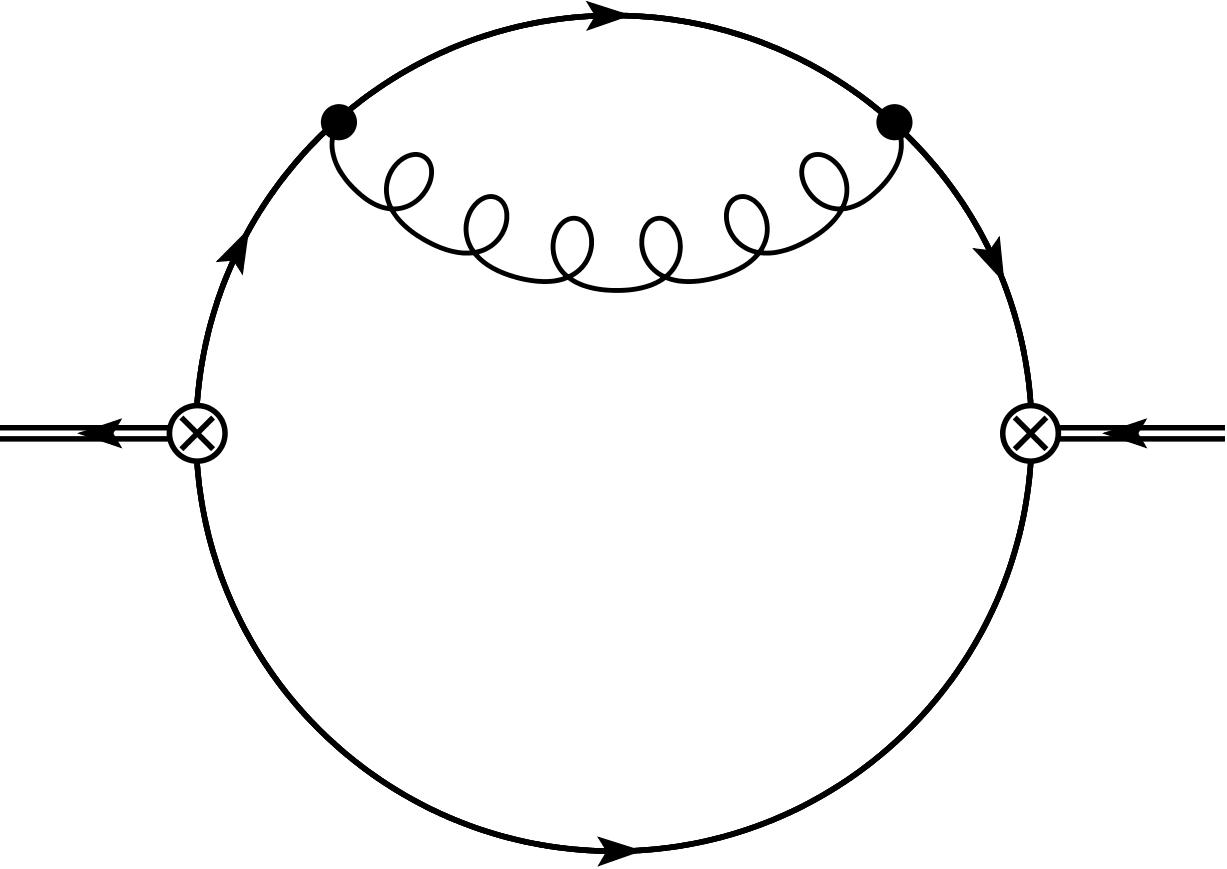} & \includegraphics[width=50mm]{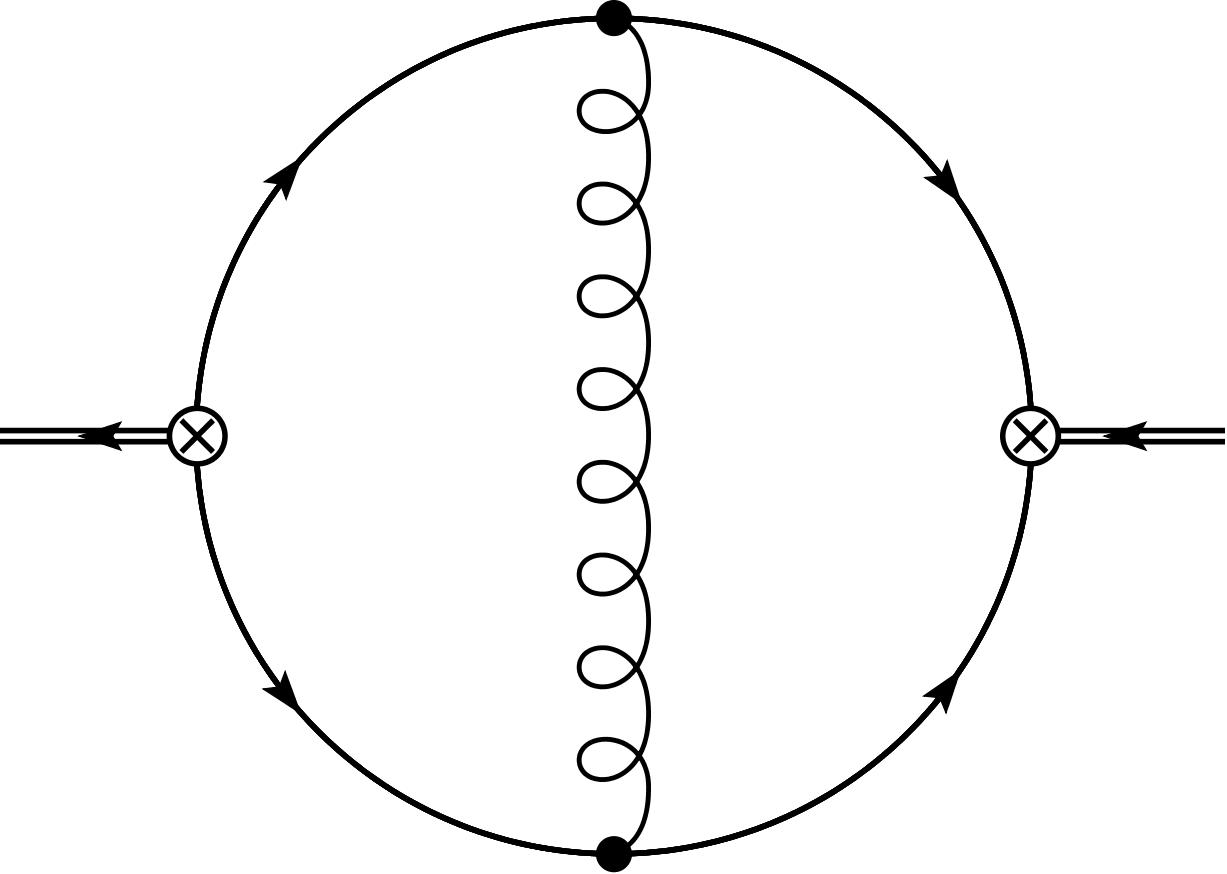}\\
Diagram \rom{1} & Diagram \rom{2} & Diagram \rom{3} \\[15pt]
\includegraphics[width=50mm]{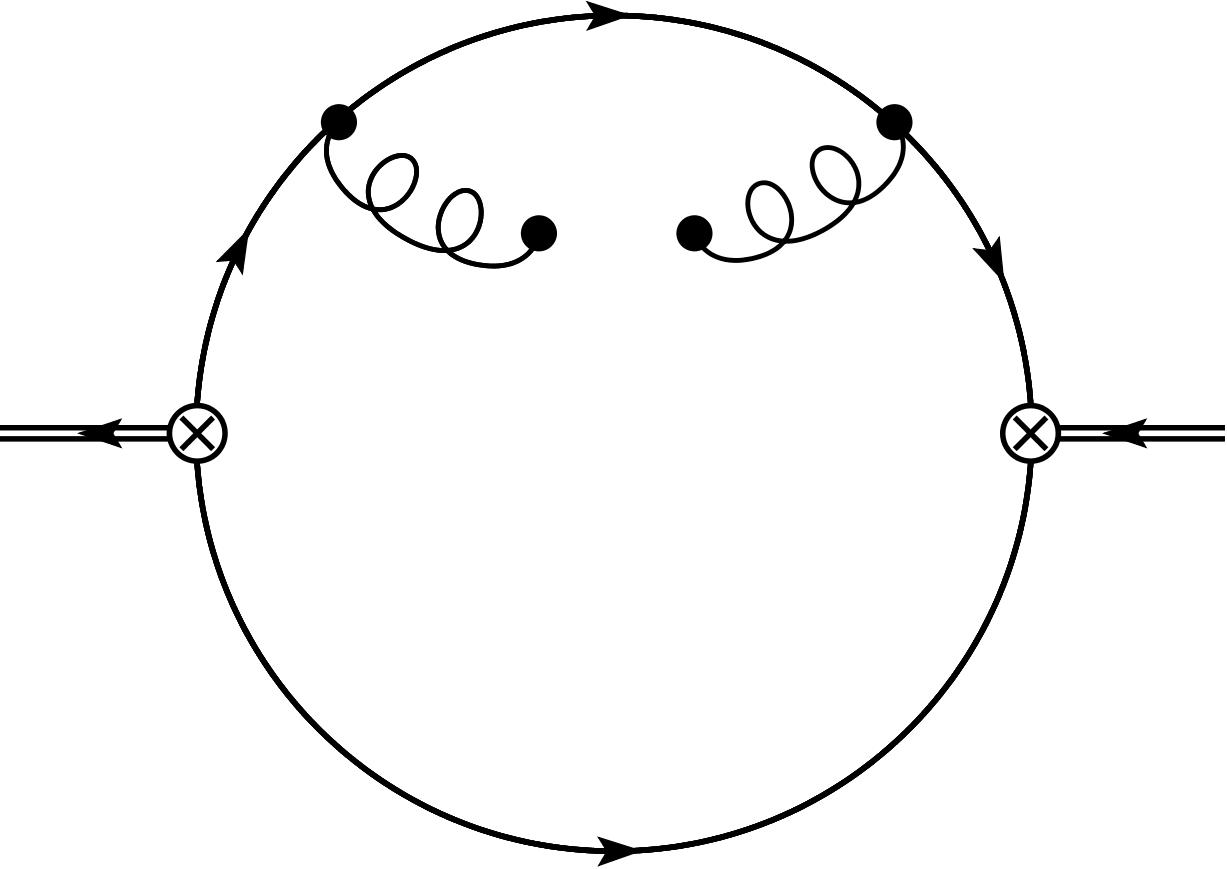} & \includegraphics[width=50mm]{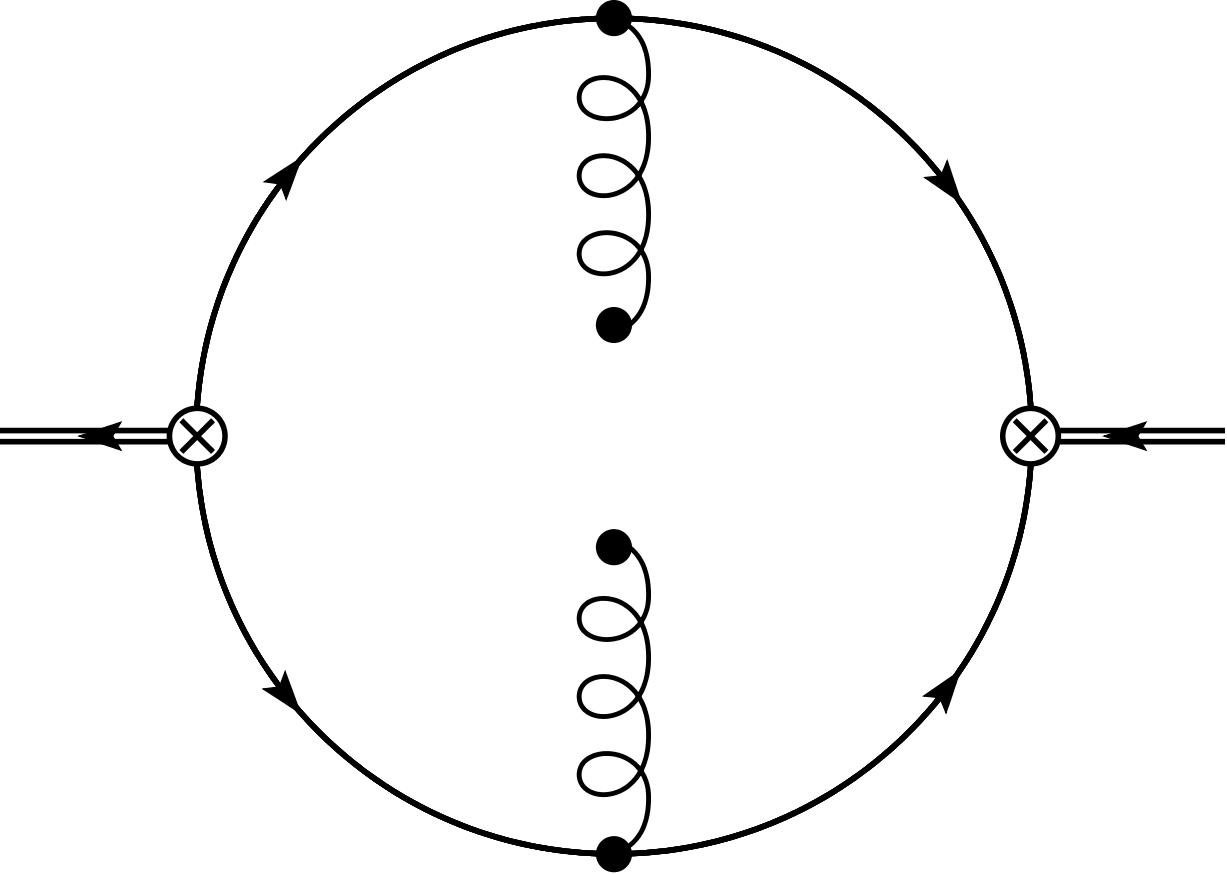} & \includegraphics[width=50mm]{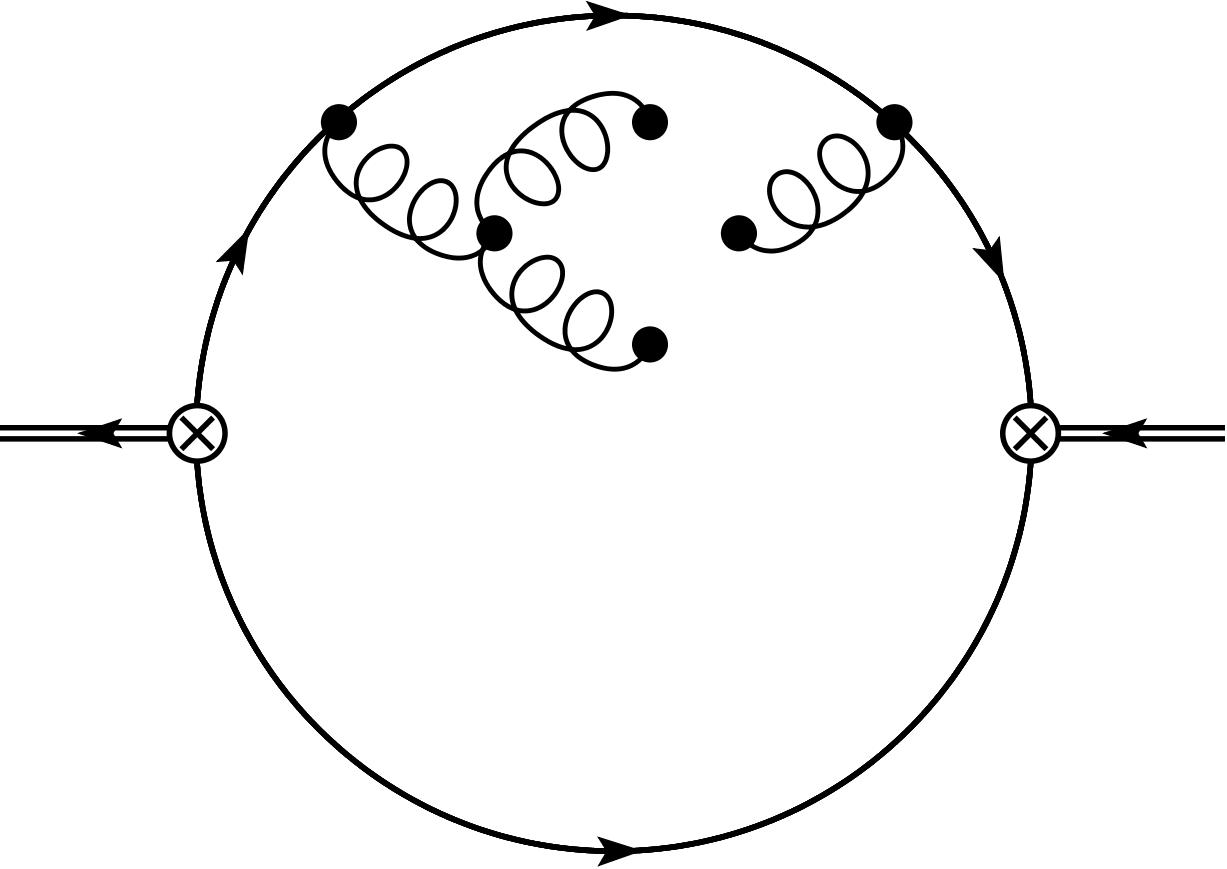}\\
Diagram \rom{4} & Diagram \rom{5} & Diagram \rom{6} \\[15pt]
\includegraphics[width=50mm]{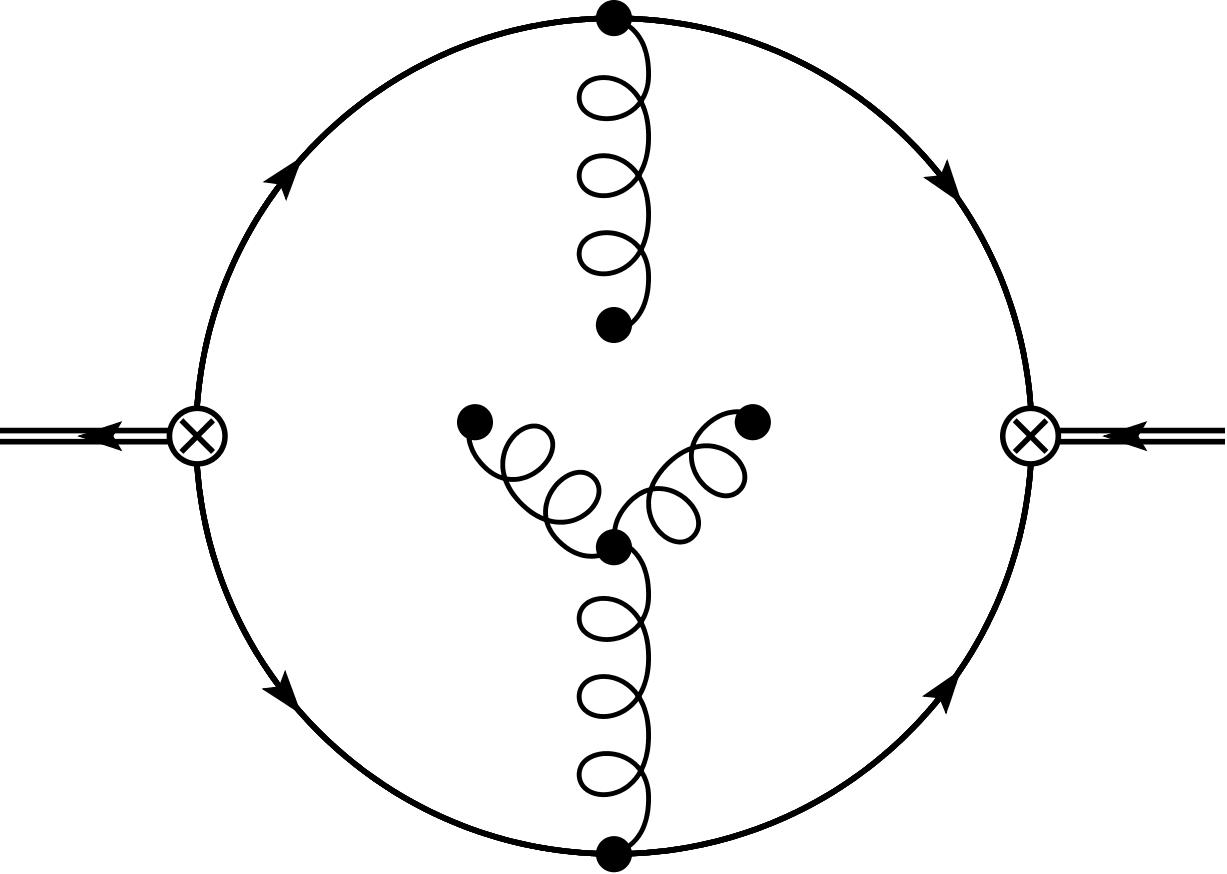} & \includegraphics[width=50mm]{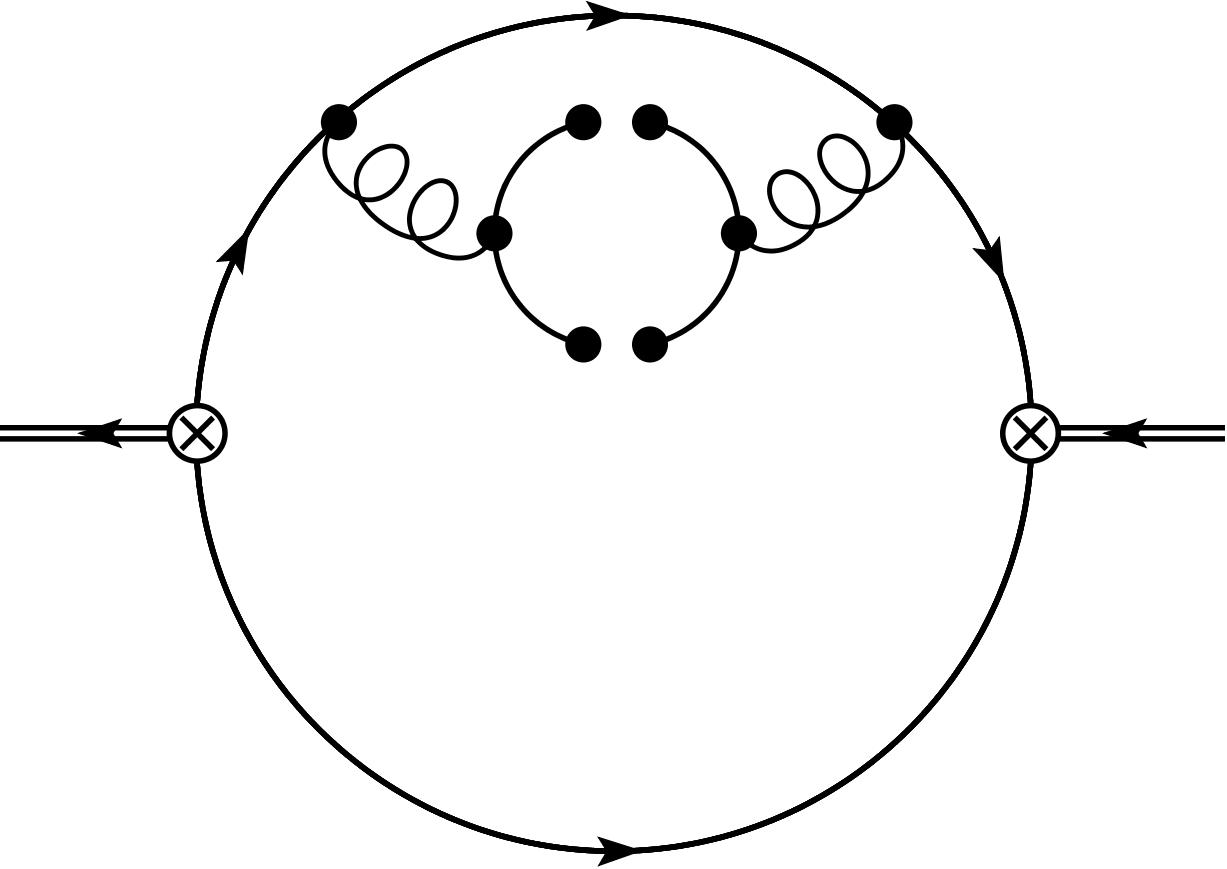} & \includegraphics[width=50mm]{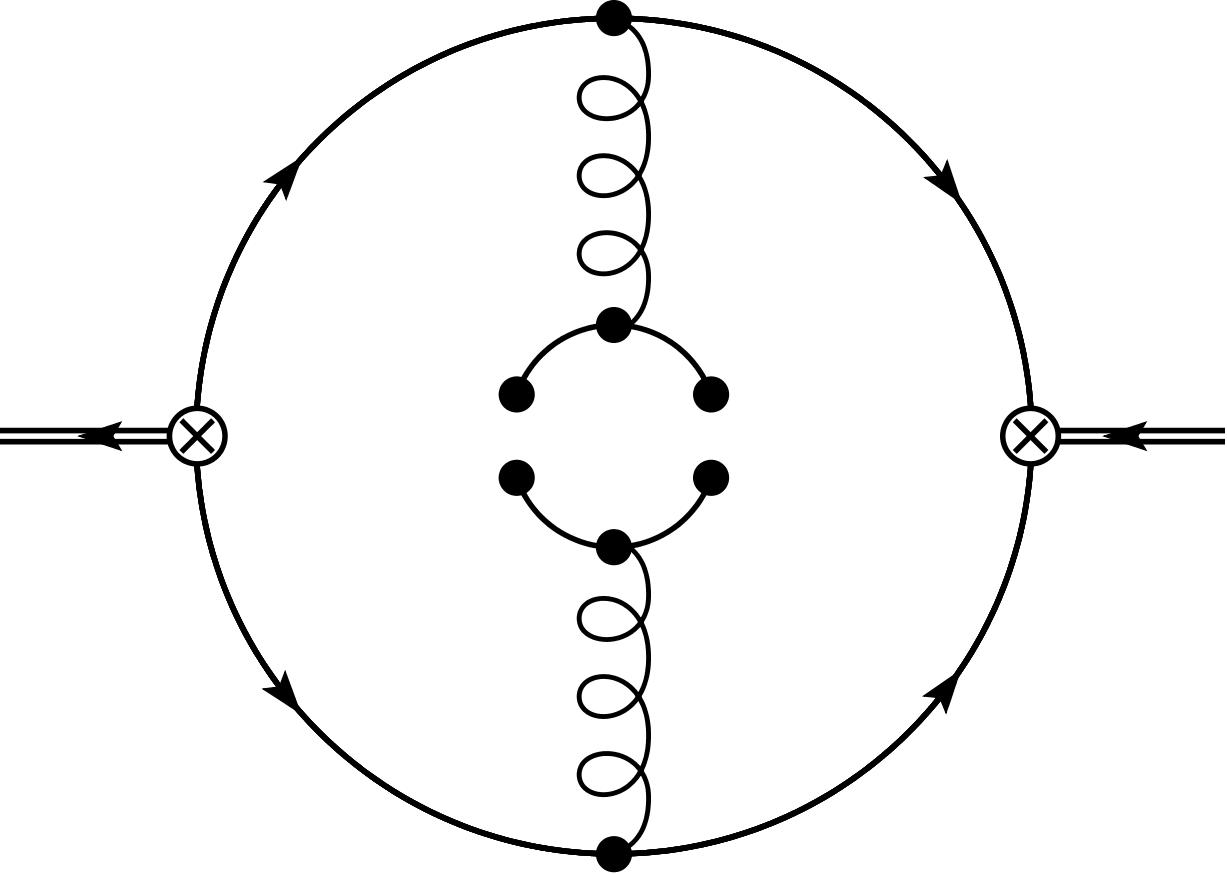}\\
Diagram \rom{7} & Diagram \rom{8} & Diagram \rom{9} \\[15pt]
\includegraphics[width=50mm]{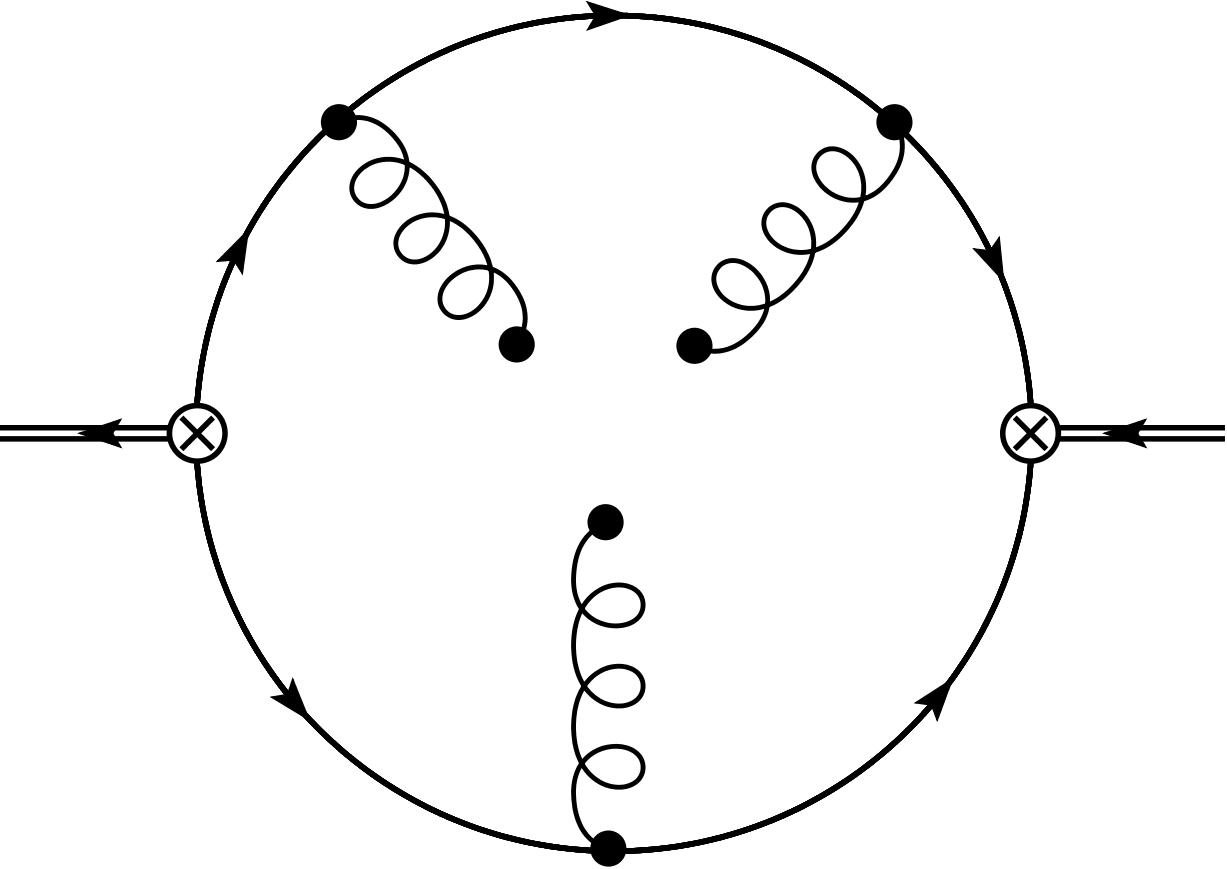} & \includegraphics[width=50mm]{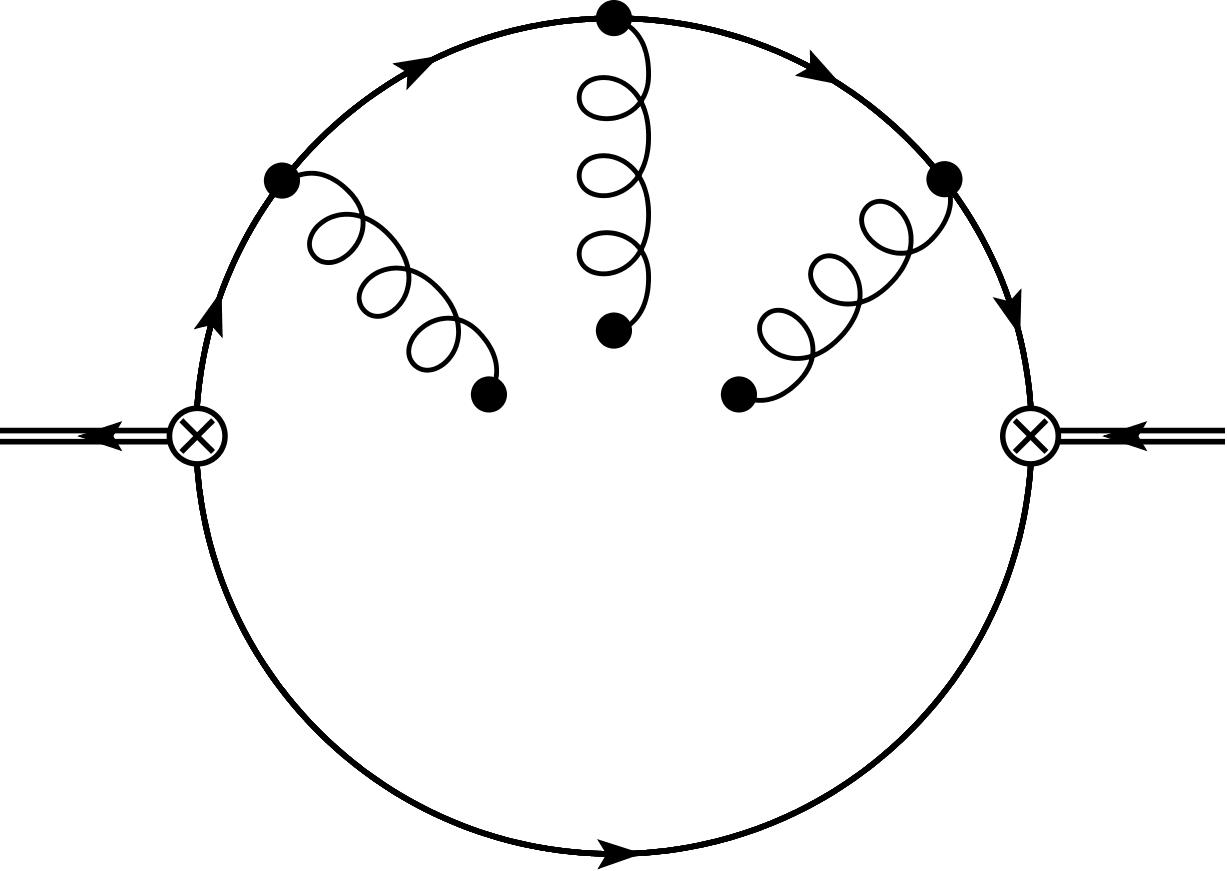} & \includegraphics[width=50mm]{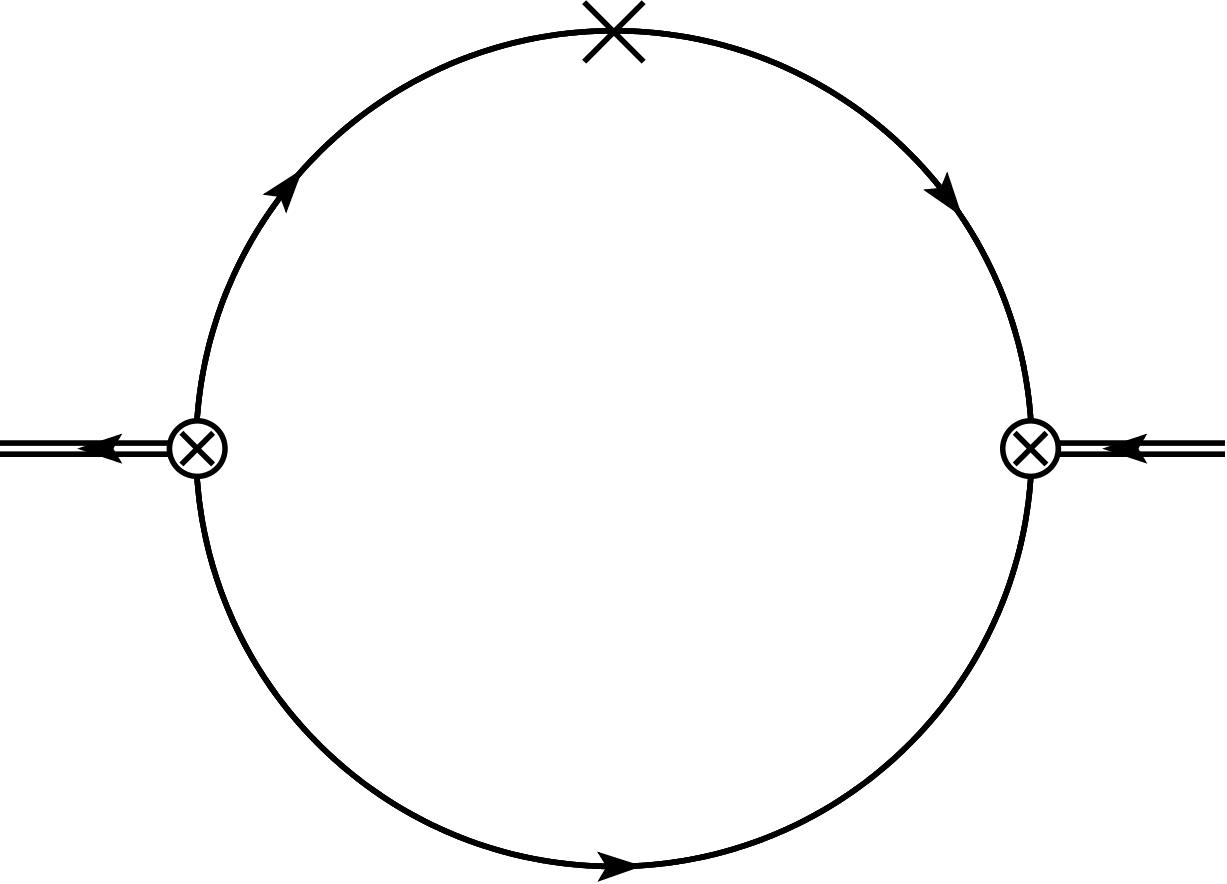}\\
Diagram \rom{10} & Diagram \rom{11} & Diagram C1 \\[15pt]
\end{tabular}
\caption{Feynman diagrams that contribute to the correlator~(\ref{Correlator}) to NLO and up to dimension-six in the QCD condensates. Diagram C1 is the counterterm diagram used to eliminate the non-local divergence in Diagram~\rom{2}. Feynman diagrams were created using JaxoDraw \cite{Binosi:2003yf}.}
\label{fig01}
\end{figure}

The diagrams computed in the simplification of~(\ref{Correlator}) are given in Figure~\ref{fig01}. 
Each diagram has a (base) multiplicity of two associated with interchanging the quark fields contracted on the top and bottom quark lines. 
Diagrams $\text{\rom{2}, \rom{4}, \rom{6}, \rom{8}, \rom{10}, \text{and} \rom{11}}$ receive an additional factor of two to account for vertical reflections.
As noted earlier,
Wilson coefficients are calculated in the Landau gauge.
Divergent integrals are handled using dimensional regularization in $D=4+2\epsilon$
dimensions at $\overline{\text{MS}}$ renormalization scale $\mu$. 
We use a dimensionally regularized $\gamma^5$ satisfying $(\gamma^5)^2=1$
and $\{\gamma^{\mu},\,\gamma^5\}=0$~\cite{ChanowitzFurmanHinchliffe1979}.
The recurrence relations of Refs.~\cite{Tarasov1996,Tarasov1997} are implemented via the Mathematica package TARCER~\cite{MertigScharf1998} resulting in expressions phrased in terms of master integrals with known solutions including those of~\cite{BoosDavydychev1991,BroadhurstFleischerTarasov1993}.

The OPE computation of $\Pi$, denoted $\Pi^{\text{OPE}}$, can be written as
\begin{equation} \label{decomposedCorrelator}
\Pi^{\text{OPE}}(q^2)=
\sum_{\text{i}=\text{\rom{1}}}^{\text{\rom{11}}}\Pi^{\text{(i)}}(q^2)
\end{equation}
where the superscript in~(\ref{decomposedCorrelator}) corresponds to the labels of the diagrams in Figure~\ref{fig01}. 
Evaluating the first term in this sum, $\Pi^{(\mathrm{\rom{1}})}$, 
expanding the result in $\epsilon$, and dropping a polynomial in $q^2$ 
(which does not contribute to the sum rules---see Section~\ref{III}), we find
\begin{equation}
\Pi^{(\mathrm{\rom{1}})}(z) = \frac{4 m^2}{3 \pi ^2} z(2 z+1) \text{H}_{1}(z)
\label{A1}
\end{equation}
where $m$ is a heavy quark mass and
\begin{equation}
  z=\frac{q^2}{4m^2}.
\end{equation}
Also,
\begin{equation}\label{hypH}
\text{H}_{1}(z)= \, _2F_1\left(1,1;\frac{5}{2};z\right),
\end{equation}
where functions of the form $_pF_q\left(\cdots;\cdots;z\right)$ are generalized hypergeometric functions, \textit{e.g.,}~\cite{AbramowitzStegun1965}. 
Note that hypergeometric functions of the form ${}_{p} F_{p-1}(\cdots;\cdots;z)$ have a branch point at
$z=1$ and a branch cut extending along the positive real semi-axis.
In evaluating $\Pi^{(\mathrm{\rom{2}})}(z)$, 
we find a nonlocal divergence which is eliminated through the inclusion of the counterterm diagram, Diagram C1, of Figure~\ref{fig01}. From this point forward, we refer to the renormalized contribution arising from the sum of Diagrams \rom{2} and C1 as $\Pi^{(\mathrm{\rom{2}})}(z)$. 
Note that, in Landau gauge, Diagram~\rom{3} does not have a nonlocal divergence corresponding to the fact that 
the (multiplicative) vector diquark renormalization constant is trivial~\cite{KleivSteeleZhangEtAl2013}.
The Mathematica package HypExp~\cite{Huber:2007dx} is used to generate the $\epsilon$-expansions of $\Pi^{(\mathrm{\rom{2}})}(z)$ and $\Pi^{(\mathrm{\rom{3}})}(z)$. 
These expansions are lengthy, and so we omit them for the sake of brevity; instead, we present the exact ($\epsilon$-dependent) results, 
\begin{equation}\label{A2}
\begin{aligned}
\Pi^{(\mathrm{\rom{2}})}&(z;\epsilon) = \frac{-\alpha_{s} m^2 \Gamma (-\epsilon ) \left(\frac{m^2}{\mu ^2}\right)^{\epsilon }}{4\pi^{3}(4\pi)^{2\epsilon}(z-1) z \epsilon  (2 \epsilon +1)} \bigg[ -12z(4\pi)^{\epsilon}(2 \epsilon +1) + \\
& m^{2 \epsilon } \left(4 z^2 \epsilon  (\epsilon +1)+z \left(8 \epsilon ^3+18 \epsilon ^2+13 \epsilon +2\right)+1\right) \Gamma (-\epsilon -1)+ \\
& 4 z \left(3 (4 \pi )^{\epsilon } (2 \epsilon +1) (2 z \epsilon +1)-m^{2 \epsilon } (2 z (\epsilon +1) (2 z+2 \epsilon -1)+2 \epsilon +1) \Gamma (1-\epsilon )\right) \text{H}_{2}(z;\epsilon) - \\
& m^{2 \epsilon } \left(-4 z^2 (\epsilon +1) (3 \epsilon +2)+z \epsilon  (6 \epsilon +5)+2 (\epsilon +1)\right) \Gamma (-\epsilon -1) \text{H}_{3}(z;\epsilon) + \\
& m^{2 \epsilon } \left(8 z^3 \epsilon  (\epsilon +1)-8 z^2 (\epsilon +1) (3 \epsilon +1)+2 z (\epsilon -1) (2 \epsilon +1)+2 \epsilon +1\right) \Gamma (-\epsilon -1) \text{H}_{4}(z;\epsilon) \bigg]
\end{aligned}
\end{equation}
\begin{equation}\label{A3}
\begin{aligned}
\Pi^{(\mathrm{\rom{3}})}&(z;\epsilon) =\frac{\alpha_{s} (\epsilon +1) m^{2\epsilon+2} \Gamma (-\epsilon -1)^2 \left(\frac{m^2}{\mu ^2}\right)^{\epsilon }}{(2\pi)^{3}(4\pi)^{2\epsilon}(z-1) z \epsilon  (4 \epsilon  (\epsilon +2)+3)^2} \bigg[ \\
& (4 \epsilon  (\epsilon +2)+3) (-z (-8 z (\epsilon +1)+\epsilon  (4 \epsilon  (\epsilon +2)+7)+2)-2 \epsilon -3) - \\
& 8 z (\epsilon +1) (2 \epsilon +1) (2 \epsilon +3) \left(4 z^2 (\epsilon +1)+z \epsilon  (2 \epsilon +3)+2 \epsilon ^2+\epsilon -1\right) \text{H}_{2}(z;\epsilon) - \\
& 4 z (\epsilon +1) (2 \epsilon +1) (4 \epsilon  (\epsilon +2)+3) \left(1-2 z (\epsilon +1) \left(2 (z-1) \epsilon ^2+z \epsilon +2 z+\epsilon \right)\right) \text{H}_{2}(z;\epsilon)^{2} + \\
& (2 \epsilon +3) (4 \epsilon  (\epsilon +2)+3) \left(-8 z^2 (\epsilon +1)^2+z \epsilon  (2 \epsilon +1)+2 (\epsilon +1)\right) \text{H}_{3}(z;\epsilon) - \\
& (4 \epsilon  (\epsilon +2)+3) \Big(-16 z^3 \begin{aligned}[t] &(\epsilon +1)-8 z^2 (\epsilon +1) (\epsilon  (2 \epsilon +7)+2)- \\
& 2 z (\epsilon  (4 \epsilon  (\epsilon +2)+5)-1)+4 \epsilon  (\epsilon +2)+3 \Big) \text{H}_{4}(z;\epsilon) \bigg],
\end{aligned}
\end{aligned}
\end{equation}
where
\begin{gather}
\text{H}_{2}(z;\epsilon)= \, _2F_1\left(1,-\epsilon ;\frac{3}{2};z\right) \\
\text{H}_{3}(z;\epsilon)= \, _3F_2\left(1,-2 \epsilon -1,-\epsilon ;\frac{1}{2}-\epsilon ,\epsilon +2;z\right) \\
\text{H}_{4}(z;\epsilon)= \, _3F_2\left(1,-2 \epsilon ,-\epsilon ;\frac{1}{2}-\epsilon ,\epsilon +2;z\right).
\end{gather}
The $\epsilon$-expanded results for the remaining terms in~(\ref{decomposedCorrelator}) can be written more concisely and are given by
\begin{gather}
\Pi^{(\mathrm{\rom{4}})}(z) = \frac{-3 \left(8 z^2-17 z+6\right) + \left(2 z^2-11 z+6\right) \text{H}_{1}(z)}{288 \pi  m^2 (z-1)^3}\big\langle\alpha G^{2} \big\rangle
\label{A4}\\
\Pi^{(\mathrm{\rom{5}})}(z) = \frac{12 z-15 - (2 z-3) \text{H}_{1}(z)}{576 \pi  m^2 (z-1)^2}\big\langle\alpha G^{2} \big\rangle
\label{A5}\\
\begin{aligned}
\Pi^{(\mathrm{\rom{6}})}(z) = \frac{\big\langle g^{3} G^{3} \big\rangle}{92160 \pi ^2 m^4 (z-1)^5 z}\Big( 416 z^5 & -1888 z^4+3078 z^3-1836 z^2+90 z+35+ \\
&5 \left(8 z^5-36 z^4+42 z^3-20 z^2+20 z-7\right) \text{H}_{1}(z)\Big)
\end{aligned}
\label{A6}\\
\Pi^{(\mathrm{\rom{7}})}(z) = \frac{32 z^3-89 z^2+19 z+8 - \left(12 z^4-66 z^3+73 z^2-37 z+8\right) \text{H}_{1}(z)}{55296 \pi ^2 m^4 (z-1)^4 z}\big\langle g^{3} G^{3} \big\rangle
\label{A7}\\
\begin{aligned}
\Pi^{(\mathrm{\rom{8}})}(z) = \frac{\alpha_{s}^{2} \big\langle \overline{q}q \big\rangle^2}{4860 m^4 (z-1)^5 z}
\Big( 3 & \left(576 z^5-2608 z^4+4458 z^3-3316 z^2+765 z+20\right)- \\
& 5 \left(8 z^5-36 z^4+66 z^3-74 z^2+3 z+12\right) \text{H}_{1}(z) \Big)
\end{aligned}
\label{A8}\\
\Pi^{(\mathrm{\rom{9}})}(z) = \frac{160 z^3-478 z^2+386 z+7 - \left(56 z^4-196 z^3+226 z^2-68 z+7\right) \text{H}_{1}(z)}{1944 m^4 (z-1)^4 z} \alpha_{s}^{2} \big\langle \overline{q}q \big\rangle^2
\label{A9}\\
\Pi^{(\mathrm{\rom{10}})}(z) = \frac{-8 z^3+19 z^2+7 z-3 + \left(4 z^4-22 z^3+23 z^2-13 z+3\right) \text{H}_{1}(z)}{55296 \pi ^2 m^4 (z-1)^4 z} \big\langle g^{3} G^{3} \big\rangle
\label{A10}\\
\Pi^{(\mathrm{\rom{11}})}(z) = \frac{-3 \left(16 z^4-56 z^3+57 z^2-z-1\right) - \left(4 z^4-14 z^3-5 z^2-3 z+3\right) \text{H}_{1}(z)}{27648 \pi ^2 m^4 (z-1)^4 z} \big\langle g^{3} G^{3} \big\rangle.
\label{A11}
\end{gather}
Finally, substituting~(\ref{A1}),~(\ref{A2}),~(\ref{A3}) and (\ref{A4})--(\ref{A11})
into~(\ref{decomposedCorrelator}) gives us $\Pi^{\text{OPE}}$.

Renormalization-group improvement requires that the strong coupling and quark mass be replaced by their corresponding running quantities evaluated at renormalization scale $\mu$~\cite{Narison:1981ts}. 
At one-loop in the $\overline{\text{MS}}$ renormalization scheme, 
for $cc$ diquarks, we have
\begin{gather}
  \alpha_s\rightarrow \alpha_s(\mu) = \frac{\alpha_s(M_{\tau})}{1 + \frac{25 \alpha_s(M_{\tau})}%
  {12\pi}\log\!{\Big(\frac{\mu^2}{M_{\tau}^2}\Big)}}
  \\ 
  m\rightarrow m_{c}(\mu) = \overline{m}_{c}\bigg(\frac{\alpha_s(\mu)}
  {\alpha_s(\overline{m}_{c})}\bigg)^{12/25}
\end{gather}
and for $bb$ diquarks,
\begin{gather}
  \alpha_s\rightarrow \alpha_s(\mu) = \frac{\alpha_s(M_Z)}{1 + \frac{23 \alpha_s(M_Z)}%
  {12\pi}\log\!{\Big(\frac{\mu^2}{M_Z^2}\Big)}}
  \\ 
  m\rightarrow m_{b}(\mu) = \overline{m}_{b}\bigg(\frac{\alpha_s(\mu)}
  {\alpha_s(\overline{m}_{b})}\bigg)^{12/23},
\end{gather}
where~\cite{Olive:2016xmw}
\begin{gather}
  \alpha_s(M_{\tau})=0.330\pm0.014 \label{a1} \\
  \alpha_s(M_{Z})=0.1185\pm0.0006 \label{a2} \\
  \overline{m}_c=(1.275\pm0.025)\ \text{GeV} \label{a3} \\
  \overline{m}_b=(4.18\pm 0.03)\ \text{GeV} \label{a4} .
\end{gather}
For $cc$ diquarks, $\mu \to \overline{m}_c$ and for $bb$ diquarks, $\mu \to \overline{m}_b$.
Finally, the following values are used for the gluon and quark
condensates~\cite{Launer:1983ib,Narison2010,ChenKleivSteeleEtAl2013}:
\begin{gather}
  \glueFourD=(0.075\pm0.02)\ \text{GeV}^4\label{glueFourDValue}\\
  \glueSixD=((8.2\pm1.0)\ \text{GeV}^2)\label{glueSixDValue}\glueFourD\\
 \big\langle \overline{q}q \big\rangle = -(0.23 \pm 0.03)^3\ \text{GeV}^3\label{quarkThreeDValue}.
\end{gather}

\section{QCD Laplace Sum-Rules, Analysis, and Results}\label{III}

We now proceed with the QCD LSRs analysis of axial vector $cc$ and $bb$ diquarks. 
Laplace sum-rules analysis techniques were originally introduced in~\cite{Shifman:1978bx,Shifman:1978by}. 
Subsequently, the LSRs methodology was reviewed in~\cite{Reinders:1984sr,Narison:2002pw}. 

The function $\Pi(q^2)$ of~(\ref{Correlator}) satisfies a dispersion relation
\begin{equation}\label{dispersion}
  \Pi(q^2) = q^4\int_{t_0}^{\infty}\!\frac{\frac{1}{\pi}\Im\Pi(t)}{t^2(t-q^2)}\,\mathrm{d}t
  + \cdots
\end{equation}
for $q^2<0$.
In~(\ref{dispersion}), $t_0$ is an effective threshold and $\cdots$ represents a polynomial in $q^2$.
On the left-hand side of~(\ref{dispersion}), $\Pi$ is identified with $\Pi^{\text{OPE}}$ 
computed in Section~\ref{II}.  On the right-hand side of~(\ref{dispersion}),
we express $\Im\Pi(t)$, \ie\ the spectral function, 
using a single narrow resonance plus continuum model,
\begin{equation}\label{res_plus_cont}
  \frac{1}{\pi}\Im\Pi(t) = 2h_+^2\, \delta(t-M^2) + \frac{1}{\pi}\Im\Pi^{\text{OPE}}(t)\theta(t-s_0),   
\end{equation}
where $M$ is the diquark constituent mass and 
$h_+$ is the diquark coupling defined by
\begin{equation}\label{efective_coupling}
  \langle \Omega\vert j_{\mu,\alpha} \vert (cc)_\beta,1^+\rangle
  =\sqrt{\frac{2}{3}}\,\delta_{\alpha\beta}\epsilon_\mu h_+,
\end{equation}
which aligns with the notation of Ref.~\cite{Jamin:1989hh}.
Also, $\theta(t)$ is a Heaviside step function and $s_0$ is the continuum  threshold parameter.
Constituent diquark masses are key input parameters of Type~I \& II diquark-antidiquark 
models of tetraquarks~\cite{Maiani:2004vq,Maiani:2014aja} (see Section~\ref{IV}).
However, as the couplings are not parameters of Type~I \& II tetraquark models, 
we eliminate them by working with ratios of LSRs (\eg\ see~(\ref{mass})).
Though not relevant for our purposes here, we note that knowledge of the coupling $h_+$ for light diquarks  allows  estimation of  baryon matrix elements of the effective weak Hamiltonian~\cite{Dosch1989, Jamin:1989hh}. 

As discussed in Ref.~\cite{Dosch1989}, the duality relation \eqref{res_plus_cont}
for diquarks is more subtle than for hadrons because diquarks are constituent degrees of freedom rather than hadron states.  Ref.~\cite{Dosch1989} argues that, similar to constituent quarks, the diquark mass and coupling should be regarded as effective  
quantities which describe the correlator at intermediate scales. 
Above the threshold $s_0$,
the diquark loses its meaning as a constituent degree of freedom, and the correlator is dominated by the parton-level quark description (see Diagram~I in Fig.~\ref{fig01}).   
In the context of lattice QCD, the 
coupling $h_+$ is proportional to the signal strength, and Ref.~\cite{Hess:1998sd} finds a remarkably clean exponential decay indicative of a single narrow resonance below the lattice cutoff $1/a^2$. In~(\ref{res_plus_cont}), $s_0$ is analogous to the lattice cutoff $1/a^2$. Thus, in the light quark sector studied in~\cite{Hess:1998sd}, 
there exists direct lattice QCD evidence supporting the spectral decomposition~(\ref{res_plus_cont}).

\begin{figure}
\centering
\includegraphics[scale=0.35]{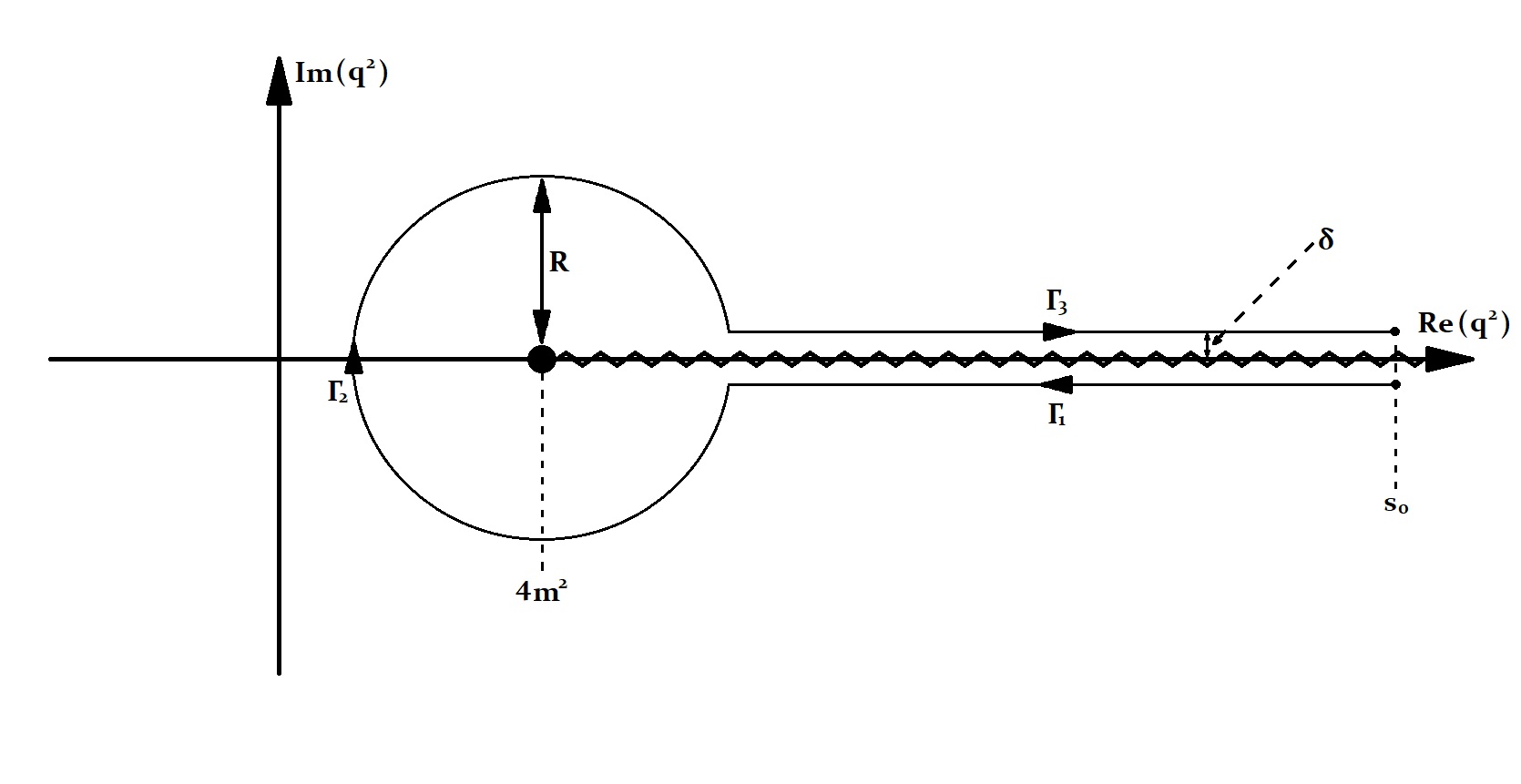}
\caption{\label{keyhole} The contour of integration used in the evaluation of the 
LSRs~(\ref{explicitLSR}). We use $\delta = 10^{-12}\ \text{GeV}^2$ and $\text{R}=2m^2$ generally in the calculation of~(\ref{explicitLSR}) however other values and contour shapes were tested to verify that the code was producing contour invariant results as it must.}
\label{Contour}
\end{figure}

Laplace sum-rules are obtained by Borel transforming~(\ref{dispersion}) weighted by powers of 
$Q^2$ (see~\cite{Shifman:1978bx,Shifman:1978by} as well as,
\textit{e.g.,}~\cite{ReindersRubinsteinYazaki1984,narisonbook:2004}).
For a function such as $\Pi^{\text{OPE}}$ computed in Section~\ref{II}, details on how to evaluate
the Borel transform can be found in~\cite{Palameta:2017ols,Palameta:2018yce} for instance.
We find 
\begin{align}
    \mathcal{R}_k(\tau) &\equiv \frac{1}{2\pi i}\int_{\Gamma} (q^2)^k e^{-q^2\tau}\Pi^{\text{OPE}}(q^2)\,\mathrm{d}q^2
    + \int_{s_0}^{\infty}\! t^k e^{-t\tau}\frac{1}{\pi}\Im\Pi^{\text{OPE}}(t)\,\mathrm{d}t
    \label{unsubtractedLSROPE}\\
    \implies \mathcal{R}_k(\tau) &= 2h_{+}^2 M^{2k} e^{-M^2 \tau}
    + \int_{s_0}^{\infty}\! t^k e^{-t\tau}\frac{1}{\pi}\Im\Pi^{\text{OPE}}(t)\,\mathrm{d}t
    \label{unsubtractedLSRhadron}
\end{align}
where $\mathcal{R}_k(\tau)$ are unsubtracted LSRs of (usually non-negative) integer order $k$
evaluated at Borel scale $\tau$ and where $\Gamma$ is the integration contour depicted in Figure~\ref{keyhole}. 
Subtracting the continuum contribution, 
\begin{equation}
   \int_{s_0}^{\infty}\! t^k e^{-t\tau}\frac{1}{\pi}\Im\Pi^{\text{OPE}}(t)\,\mathrm{d}t, 
\end{equation}
from the right-hand sides of (\ref{unsubtractedLSROPE}) and (\ref{unsubtractedLSRhadron}), we find
\begin{align}
    \mathcal{R}_k(\tau,\,s_0) &\equiv \frac{1}{2\pi i}\int_{\Gamma} (q^2)^k e^{-q^2\tau}\Pi^{\text{OPE}}(q^2)\,\mathrm{d}q^2
    \label{subedLSR_OPE}\\
    \implies \mathcal{R}_k(\tau,\,s_0) &= 2h_{+}^2 M^{2k} e^{-M^2 \tau}
    \label{subedLSR}
\end{align}
where $\mathcal{R}_k(\tau,\,s_0)$ are (continuum-)subtracted LSRs. 
In~(\ref{subedLSR_OPE}), explicitly parametrizing each $\Gamma_i$ of $\Gamma$, we have
\begin{equation}\label{explicitLSR}
\begin{aligned}
\mathcal{R}_{k}(\tau,s_0)\equiv\frac{1}{2\pi i} \bigg[
&\int_{s_0}^{4m^2- \; \sqrt[]{\text{R}^2-\delta^2}} (t-\delta i)^k e^{-(t-\delta i) \tau}
\Pi^{\text{OPE}}(t-\delta i) \dif{t} + \\
& \int_{2\pi-\sin^{-1}(\delta/\text{R})}^{\sin^{-1}(\delta/\text{R})} (4m^2+\text{R} e^{\theta i})^k e^{-(4m^2+\text{R} e^{\theta i}) \tau} \text{R} i e^{\theta i} \Pi^{\text{OPE}}(4m^2+\text{R} e^{\theta i}) \dif{\theta} + \\
&\int^{s_0}_{4m^2- \; \sqrt[]{\text{R}^2-\delta^2}} (t+\delta i)^k e^{-(t+\delta i) \tau}
\Pi^{\text{OPE}}(t+\delta i) \dif{t} \bigg],
\end{aligned}
\end{equation}
which is then calculated numerically.
In~(\ref{explicitLSR}), $R$ is set to $2m^2$. 
Also, it is intended that $\delta\rightarrow 0^{+}$.  In practice, this can be achieved
by setting $\delta = 10^{-12}\ \text{GeV}^2$.  Finally, using~(\ref{subedLSR}), we find
\begin{equation}\label{mass}
\sqrt{
\frac{\mathcal{R}_{1}(\tau,s_0)}{\mathcal{R}_{0}(\tau,s_0)}
}
= M.
\end{equation}

To use~(\ref{mass}) to predict diquark constituent masses,
we must first select an acceptable range of $\tau$ values, \textit{i.e.,} a Borel window $\double{\tau_{\text{min}}}{\tau_{\text{max}}}$. 
To determine the Borel window, we follow the methodology outlined in~\cite{Kleiv:2013dta}. 
To generate $\tau_{\text{max}}$, we require OPE convergence of the $k=0$ LSRs 
as $s_0 \to \infty$. 
By OPE convergence, we mean that the total perturbative contribution to the LSRs (pert), 
the total 4d contribution to the LSRs (4d), and the total 6d contribution to the
LSRs (6d) must obey the inequality
\begin{equation}\label{converge}
|\text{pert}| \geq 3 \times |\text{4d}| \geq 9 \times |\text{6d}|.
\end{equation}
The lowest value of $\tau$ for which~(\ref{converge}) is violated as 
$s_0 \to \infty$ becomes $\tau_{\text{max}}$.
Additionally, $\tau_{\text{max}}$ is constrained by the requirement
\begin{equation}\label{Holder}
  \frac{\mathcal{R}_{2}(\tau,s_0)/\mathcal{R}_{1}(\tau,s_0)}{\mathcal{R}_{1}(\tau,s_0)/\mathcal{R}_{0}(\tau,s_0)} 
  \geq 1
\end{equation}
where this inequality results from requiring that individually both $\mathcal{R}_{1}(\tau,s_0)$ and $\mathcal{R}_{0}(\tau,s_0)$ satisfy the H\"older inequalities~\cite{Beckenbach1961,Berberian1965} 
as per~\cite{Kleiv:2013dta}.
For the specific LSRs being studied here, it turns out that the condition~(\ref{converge}) is more
restrictive than the condition~(\ref{Holder}).
For both diquark channels under consideration, the values of $\tau_{\text{max}}$ obtained  
are given in the last column of Table~\ref{results}.
To select $\tau_{\text{min}}$, in addition to the H\"older inequality constraint~\eqref{Holder}, we require that
\begin{equation}\label{pole_contribution}
  \frac{\mathcal{R}_{1}(\tau,s_0)/\mathcal{R}_{0}(\tau,s_0)}{\mathcal{R}_{1}(\tau,\infty)/\mathcal{R}_{0}(\tau,\infty)} \geq 0.5
\end{equation}
\ie\ that the resonance contribution to $\mathcal{R}_1/\mathcal{R}_0$ must be at least 50\%.
The highest value of $\tau$ which does not violate~\eqref{Holder}--(\ref{pole_contribution}) becomes $\tau_{\text{min}}$. 
For both diquark channels under consideration, the values of $\tau_{\text{min}}$ obtained  
are given in the second-to-last column of Table~\ref{results}.

The procedure described above for choosing a Borel window is $s_0$-dependent.
However, $s_0$ is a parameter that is predicted using the optimization procedure described below.  
As such, choosing a Borel window and predicting $s_0$ are actually handled iteratively.
Typically, the Borel window widens as $s_0$ increases. 
As such, we begin by selecting the minimum value of $s_0$ for which a Borel window exists.
The corresponding Borel window is then used to predict a new, updated $s_0$.  
This new $s_0$ is then used to update the Borel window 
which, in turn, is used to update $s_0$
and so on until both the Borel window and $s_0$ settle.  
This iterative process has been taken into account in reporting diquark constituent masses, 
continuum thresholds, and Borel windows in Table~\ref{results}.

\begin{figure}[htb]
    \centering
    \includegraphics[scale=0.8]{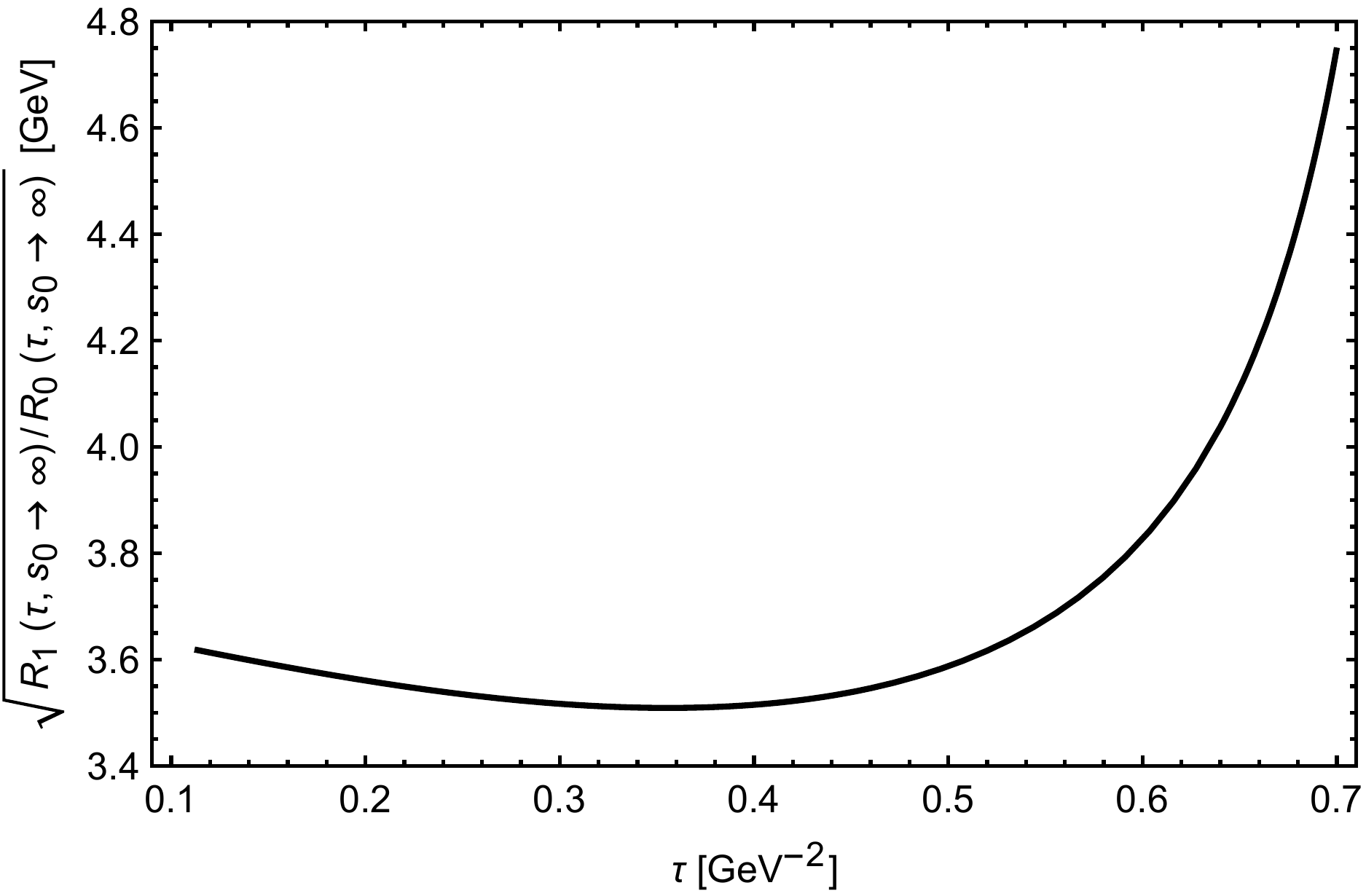}
    \caption{The left-hand side of~(\ref{mass}) at the optimized 
    continuum threshold parameter $s_0$ (see Table~\ref{results}) versus the Borel scale $\tau$
    for the $cc$ diquark.}
    \label{axialccGraph}
\end{figure}

\begin{figure}[htb]
    \centering
    \includegraphics[scale=0.8]{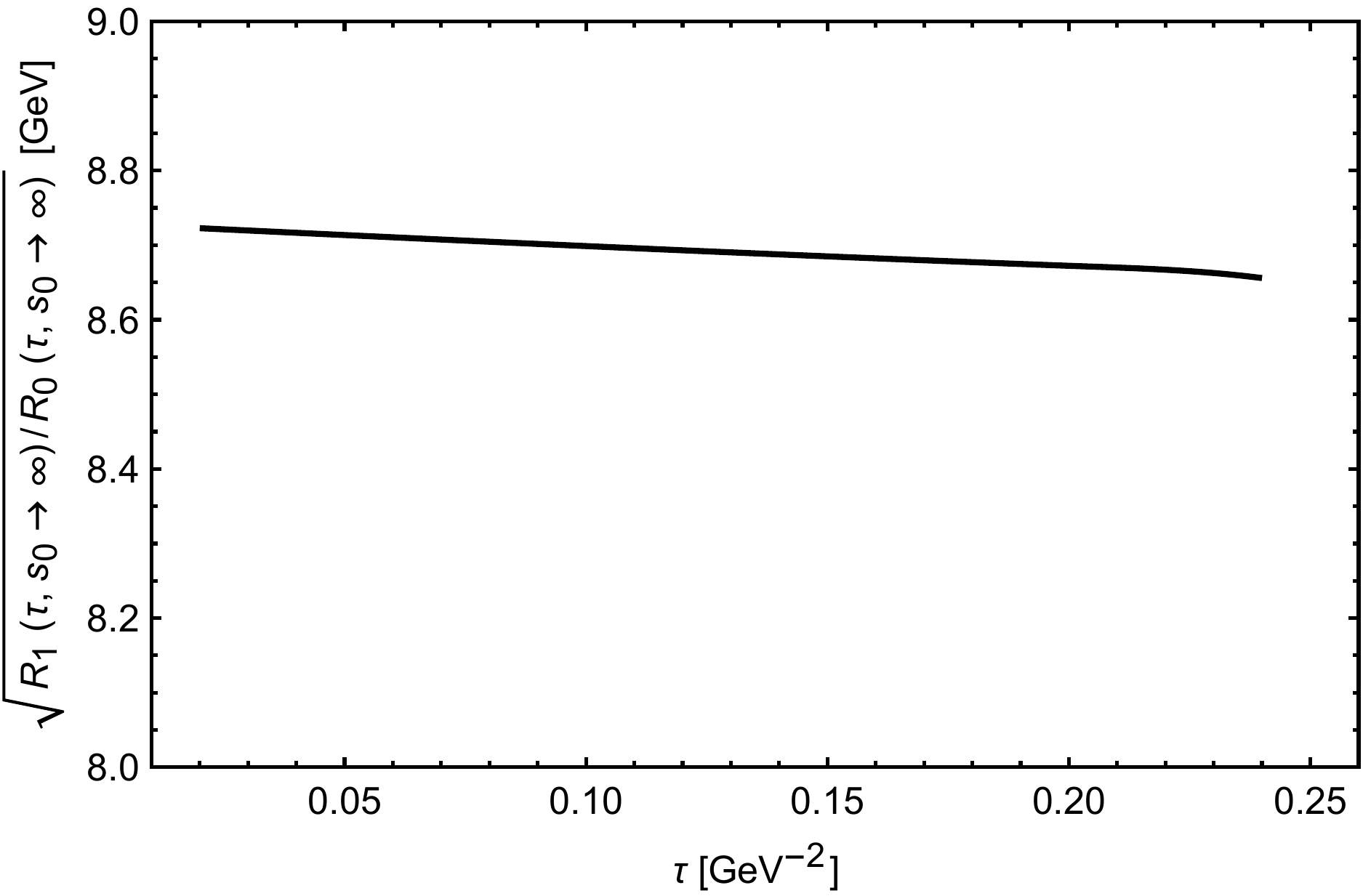}
    \caption{The left-hand side of~(\ref{mass}) at the optimized 
    continuum threshold parameter $s_0$ (see Table~\ref{results}) versus the Borel scale $\tau$
    for the $bb$ diquark.}
    \label{axialbbGraph}
\end{figure}

To predict $s_0$ and $M$, we optimize the agreement between left- and right-hand sides of~(\ref{mass})
by minimizing
%
\begin{equation}\label{chiSq}
\chi^2(s_0,\,M) = \sum_{j=0}^{20} \bigg( \frac{1}{M} \sqrt{\frac{\mathcal{R}_{1}(\tau_j,s_0)}{\mathcal{R}_{0}(\tau_j,s_0)}} -1 \bigg)^2
\end{equation}
where we have partitioned the Borel window into 20 equal length subintervals with $\{\tau_j\}_{j=0}^{20}$.
For both diquark channels under consideration, the optimized values of $s_0$ obtained  
are given in the third column of Table~\ref{results}.
As a consistency check on our methodology, we require that the optimized mass $M$
from~(\ref{chiSq}) actually yields a good fit to~(\ref{mass}) and that the left-hand side of~(\ref{mass})
exhibits $\tau$ stability~\cite{Kleiv:2013dta}, that is
%
%
\begin{equation}\label{dTau}
    \frac{\mathrm{d}}{\mathrm{d}\tau} \sqrt{\frac{\mathcal{R}_1(\tau,\,s_0)}{\mathcal{R}_0(\tau,\,s_0)}} \approx 0
\end{equation}
within the Borel window. 
And so, in Figures~\ref{axialccGraph} and~\ref{axialbbGraph}, 
we plot the left-hand side of~(\ref{mass}) at the appropriate optimized $s_0$ 
versus $\tau$ over the appropriate Borel window for both diquark channels under consideration.
For the $bb$ diquark, the optimized $M$ from~(\ref{chiSq}) does indeed yield a good fit
to~(\ref{mass}).  
Specifically, $M = 8.67$~GeV in agreement with Figure~\ref{axialbbGraph}.
Regarding condition~(\ref{dTau}), over the Borel window,
\begin{equation}
    \frac{1}{M}\,\left\lvert\Delta\!\left(\sqrt{\frac{\mathcal{R}_1(\tau,\,s_0)}%
    {\mathcal{R}_0(\tau,\,s_0)}}\right)\right\rvert
    \approx 0.001,
\end{equation}
implying that the plot in Figure~\ref{axialbbGraph} can be considered flat to an 
excellent approximation.
For the $cc$ diquarks, it is clear from Figure~\ref{axialccGraph} that the fitted value of $M$ will be biased by the rapid increase at large $\tau$ values.  
We thus use the critical point $\frac{\mathrm{d}}{\mathrm{d}\tau} \sqrt{\mathcal{R}_1/\mathcal{R}_0} =0  $ for our $cc$ diquark mass prediction, \ie\ $M = 3.51$~GeV.
For both diquark channels under consideration, predicted diquark  constituent masses $M$
are given in the second column of Table~\ref{results}.
The theoretical uncertainties associated with the mass predictions take into account the uncertainties arising from the strong coupling and mass parameters~(\ref{a1})--(\ref{a4}) as well as those  associated with the QCD condensate values~(\ref{glueFourDValue})--(\ref{quarkThreeDValue}). 
The dominant theoretical uncertainty is associated with the quark masses.

\begin{figure}[htb]
    \centering
    \includegraphics[scale=0.8]{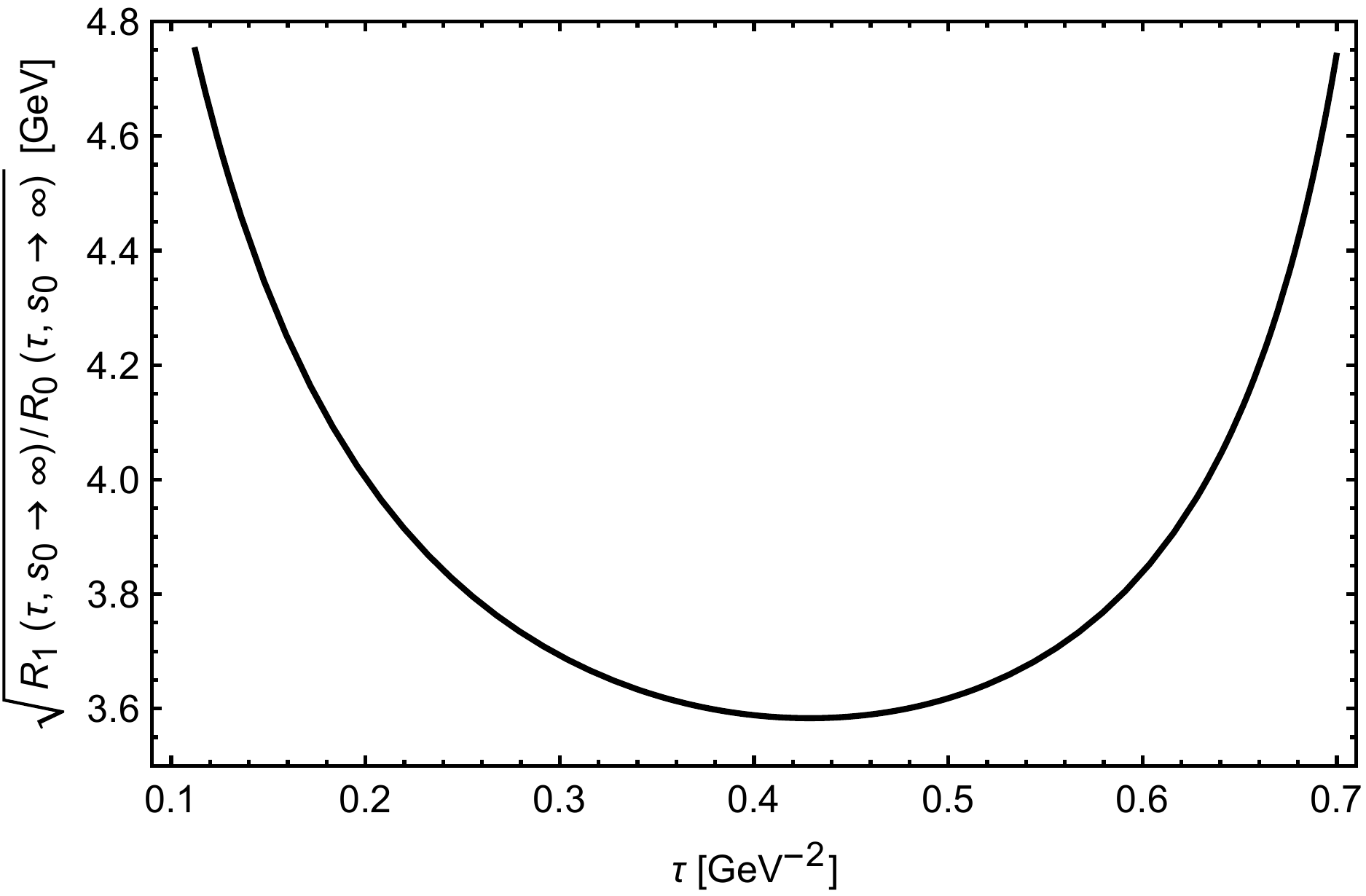}
    \caption{The left-hand side of~(\ref{mass}) 
    as the continuum threshold parameter $s_0\rightarrow\infty$ 
    versus the Borel scale $\tau$ for the $cc$ diquark.}
    \label{CCInf}
\end{figure}

\begin{figure}[htb]
    \centering
    \includegraphics[scale=0.8]{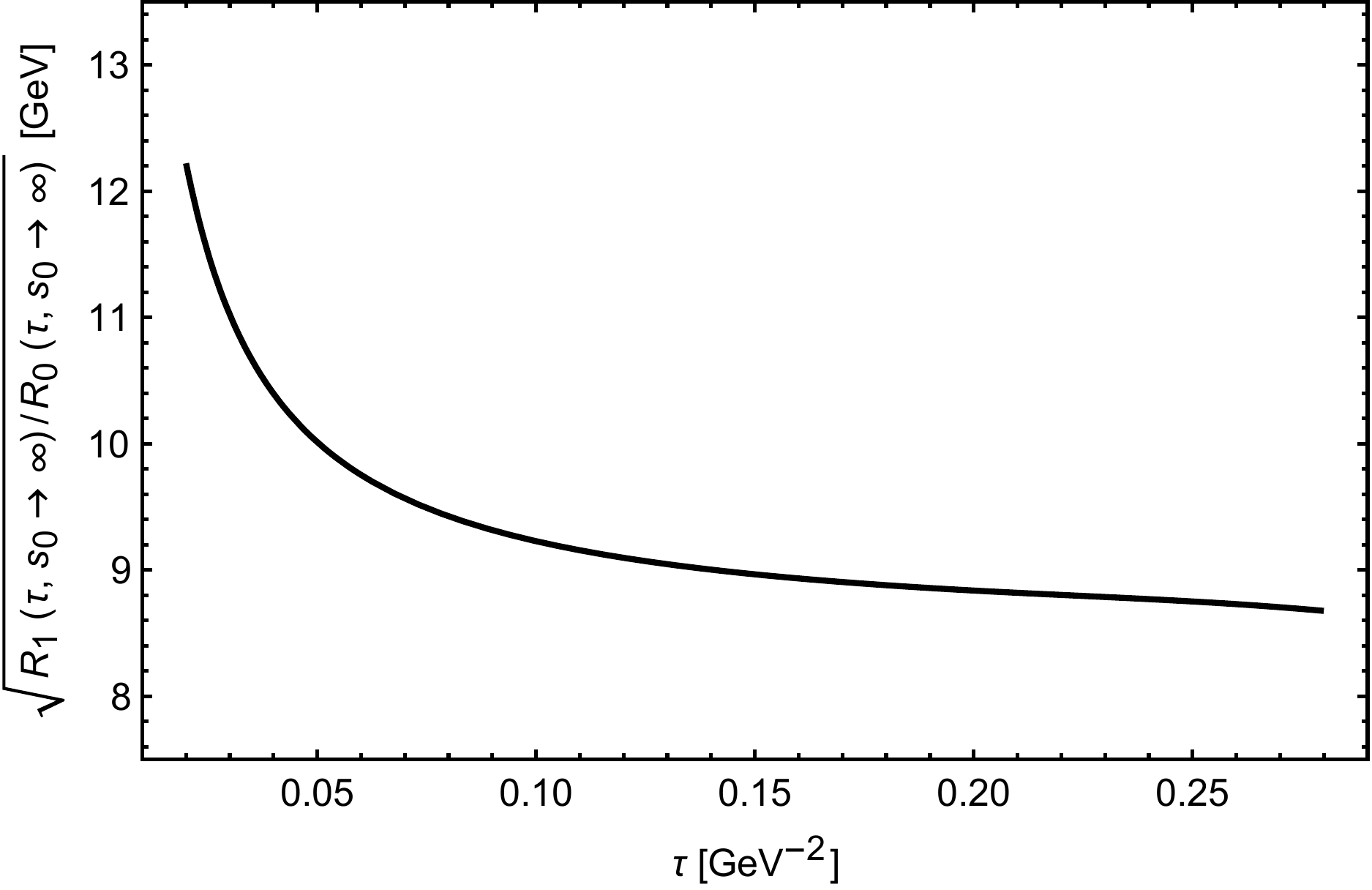}
    \caption{The left-hand side of~(\ref{mass}) 
    as the continuum threshold parameter $s_0\rightarrow\infty$ 
    versus the Borel scale $\tau$ for the $bb$ diquark.}
    \label{BBInf}
\end{figure}

In the $s_0\to\infty$ limit, the left-hand side of~\eqref{mass} 
corresponds to an upper bound on $M$ for a wide variety of resonance shapes~\cite{Harnett:2000xz}, 
allowing the sensitivity to the threshold $s_0$ and resonance model to be explored.  
As shown in Figs.~\ref{CCInf}--\ref{BBInf},
 within the Borel window $\tau<\tau_{\text{max}}$, we find $M\lesssim3.6\ \text{GeV}$ for the $cc$ case 
and $M\lesssim8.8\ \text{GeV}$ for the $bb$ case, 
remarkably close to the Table~\ref{results} predictions.


%
\begin{table}
\centering
\begin{tabular}{cccccc}
  \addlinespace
  \toprule
  $QQ$ & $M_P$ (GeV) & $s_0$ (GeV$^2$) & $\tau_{\text{min}}$ (GeV$^{-2}$) & $\tau_{\text{max}}$ (GeV$^{-2}$)\\
  \midrule
  $cc$ & $3.51 \pm 0.35$ & $17.5 \pm 3.4$ & $0.10 \pm 0.02$ & $0.71 \pm 0.07$ \\
  $bb$ & $8.67 \pm 0.69$ & $80.0 \pm 9.2$ & $0.02 \pm 0.01$ & $0.21 \pm 0.02$ \\
  \bottomrule
\end{tabular}
\caption{constituent mass predictions and sum rule parameters for axial vector $cc$ and $bb$ diquarks. The theoretical uncertainties are obtained by varying the the QCD input parameters in Eqs.~\eqref{a1}--\eqref{quarkThreeDValue}.
}
\label{results}
\end{table}
%

%


\section{Discussion }\label{IV}
 
Compared with potential model approaches~\cite{Kiselev:2002iy,Ebert:2007rn,Debastiani:2017msn}
(and others cited therein)
our $cc$ central value diquark constituent mass prediction is slightly larger 
and $bb$ is slightly smaller. 
For Bethe-Salpeter approaches~\cite{Yu:2018com}, there is closer alignment in the $cc$ constituent mass
prediction, but the $bb$ constituent mass prediction is still slightly smaller.  
However, taking into account theoretical uncertainties, we find good agreement between our QCD LSRs 
mass predictions and those of Refs.~\cite{Kiselev:2002iy,Ebert:2007rn,Debastiani:2017msn,Yu:2018com}, 
providing QCD evidence to support the study of diquark-antidiquark 
tetraquarks and doubly-heavy baryons with diquark cluster models.

Constituent diquark masses are key inputs into chromomagnetic interaction (CMI) models of
diquark-antidiquark tetraquarks.  
For example, consider the Type-II model of Ref.~\cite{Maiani:2014aja} in which colour-spin
interactions are ignored
except between the quarks (antiquarks) within the diquark (antidiquark).
This simplification assumes that the diquark and antidiquark within the tetraquark are
point-like and well-separated.
Focusing on S-wave combinations of doubly-heavy, equal mass diquarks and antidiquarks, 
the Type-II CMI Hamiltonian reduces to~\cite{Maiani:2014aja}
\begin{equation}\label{CMIHamiltonian}
    H = m_{[Q_1 Q_1]} + m_{[\bar{Q}_2 \bar{Q}_2]} 
    + 2 \kappa_{Q_1 Q_1}(\vec{S}_{Q_1}\cdot\vec{S}_{Q_1}) 
    + 2 \kappa_{\bar{Q}_2 \bar{Q}_2}(\vec{S}_{\bar{Q}_2}\cdot\vec{S}_{\bar{Q}_2})
\end{equation}
where $m_{[Q_1 Q_1]}$ and $m_{[\bar{Q}_2 \bar{Q}_2]}$ are constituent 
diquark and antidiquark masses respectively 
and where $\kappa_{Q_1 Q_1}$ and $\kappa_{\bar{Q}_2 \bar{Q}_2}$ are colour-spin interaction 
coefficients.
(Note that $\kappa_{\bar{Q} \bar{Q}}$ and $\kappa_{Q Q}$ are equal as are $m_{[QQ]}$ and 
$m_{[\bar{Q}\bar{Q}]}$.)
As the (anti-)diquarks have $J=1$, they must have $S=1$ for $L=0$
(where $J,\,L,\,S$ are the usual angular momentum quantum numbers).
Hence, the Hamiltonian~(\ref{CMIHamiltonian}) simplifies to
\begin{equation}\label{CMIHamiltonianReduced}
    H = m_{[Q_1 Q_1]} + m_{[\bar{Q}_2 \bar{Q}_2]} 
    + \frac{1}{2}\left(\kappa_{Q_1 Q_1} + \kappa_{\bar{Q}_2 \bar{Q}_2}\right).
\end{equation}
Our predictions for $m_{[cc]}$ and $m_{[bb]}$ are in Table~\ref{results}; however, the
coefficients $\kappa_{cc}$ and $\kappa_{bb}$ are not known.
In~\cite{Maiani:2014aja}, the $X(3870)$, $Z(3900)$, and $Z(4020)$
resonances were interpreted as Type-II diquark-antidiquark tetraquarks and were
used to predict $\kappa_{cq}=67\ \text{MeV}$ where $q$ is a light quark.  
As the $\kappa$ coefficients are expected to decrease with increasing
quark masses~\cite{Maiani:2004vq}, we assume here that 
\begin{equation}\label{kappas}
0 < \kappa_{cc},\,\kappa_{bc},\,\kappa_{bb} < 67\ \text{MeV}.
\end{equation}
The absolute uncertainties in our diquark  constituent mass predictions in Table~\ref{results} are
significantly larger than 67~MeV, and so, as a first approximation,
we simply ignore the $\kappa$ contributions to~(\ref{CMIHamiltonianReduced}).
Therefore, within the Type-II diquark-antidiquark model,
we predict $J^{P}\in\{0^+,\,1^+,\,2^+\}$ tetraquark masses of
7.0~GeV for $[cc][\bar{c}\bar{c}]$,
12.2~GeV for $[cc][\bar{b}\bar{b}]$, and
17.3~GeV for $[bb][\bar{b}\bar{b}]$.
The relative uncertainty in these mass predictions is roughly 10\%.
Furthermore, note that the $[cc][\bar{c}\bar{c}]$ and $[bb][\bar{b}\bar{b}]$ tetraquarks are 
charge conjugation eigenstates where $C=+$ for $J=0,\,2$ and $C=-$ for
$J=1$~\cite{Wang:2019rdo,Liu:2019zuc}.
The $[cc][\bar{b}\bar{b}]$ tetraquarks are not charge conjugation eigenstates.

Regarding $[cc][\bar{c}\bar{c}]$ tetraquarks, taking into account 10\% theoretical
uncertainty, our Type-II model mass predictions are in
reasonable agreement with those of~\cite{Wang:2019rdo,Liu:2019zuc,Chen:2016jxd},
although our central values are higher.
However, our results are much higher than those of~\cite{Wang:2017jtz}.
Furthermore, our tetraquark mass predictions are above both the 
$\eta_c(1S)$-$\eta_c(1S)$ and $J/\psi$-$J/\psi$ thresholds indicating that the corresponding 
decay modes should be accessible as fall-apart decays.

Regarding $[cc][\bar{b}\bar{b}]$ tetraquarks, again factoring in 10\% uncertainty,
our Type-II model mass predictions are in reasonable agreement with those of~\cite{Wang:2019rdo,Liu:2019zuc},
although our central values are lower.
With an electric charge of $+2$, two charm quarks, and two bottom antiquarks,
such a state would be easy to identify through its decay products, and could
not be misinterpreted as a conventional meson.
Unfortunately, within theoretical uncertainty, we are unable to say whether or
not our tetraquark mass predictions lie above or below the $B_c^+$-$B_c^+$ threshold.

Regarding $[bb][\bar{b}\bar{b}]$ tetraquarks, taking into account theoretical
uncertainty, our Type-II model mass predictions are in reasonable agreement with those
of~\cite{Chen:2016jxd} although our central values are lower.
Our results are about 10\% lower than those of~\cite{Wang:2017jtz,Anwar:2017toa},
and are much lower than those of~\cite{ Wang:2019rdo,Liu:2019zuc}.
Tetraquarks with $bb\bar{b}\bar{b}$ quark composition (so-called ``beauty-full'' tetraquarks)
have attracted considerable attention recently due to the possibility that some
might have masses below the $\Upsilon(1S)$-$\Upsilon(1S)$ threshold
and perhaps even the $\eta_b(1S)$-$\eta_b(1S)$ threshold.
For $bb\bar{b}\bar{b}$ tetraquarks with masses below the $\eta_b(1S)$-$\eta_b(1S)$ threshold,
fall-apart modes would be inaccessible and decays would instead proceed through
OZI-suppressed processes.
Central values of our Type-II diquark-antidiquark 
mass estimates put the $0^{++}$, $1^{+-}$, and $2^{++}$ states
about 9\% below the $\Upsilon(1S)$-$\Upsilon(1S)$ threshold
and about 7\% below the $\eta_b(1S)$-$\eta_b(1S)$ threshold.

In summary, we have used QCD  LSRs 
to predict the axial vector doubly-heavy $cc$ and $bb$ diquark constituent masses.
Our results are summarized in Table~\ref{results}. 
These results were obtained from a calculation of the diquark correlation function at NLO in 
perturbation theory
and to LO in the 4d and 6d gluon condensates as well as the 6d quark condensate. 
That the LSRs analyses stabilized in both the double charm and double bottom diquark
channels provides QCD-based evidence for the existence of these structures.
Within the Type-II diquark-antidiquark tetraquark model of Ref.~\cite{Maiani:2014aja}, 
we predict, with an uncertainty of roughly 10\%,
$0^{++}$, $1^{+-}$, and $2^{++}$ $[cc][\bar{c}\bar{c}]$ tetraquarks of mass 7.0~GeV;
$0^{+}$, $1^{+}$, and $2^{+}$ $[cc][\bar{b}\bar{b}]$ tetraquarks of mass 12.2~GeV; and
$0^{++}$, $1^{+-}$, and $2^{++}$ $[bb][\bar{b}\bar{b}]$ tetraquarks of mass 17.3~GeV.
Central values of our $[bb][\bar{b}\bar{b}]$ tetraquark mass predictions are well below 
the $\Upsilon(1S)$-$\Upsilon(1S)$ and $\eta_b(1S)$-$\eta_b(1S)$ thresholds, 
providing support for the possibility that fall-apart decay modes are inaccessible to 
some $bb\bar{b}\bar{b}$ tetraquarks.

\clearpage
\section*{Acknowledgements}
We are grateful for financial support from the National Sciences and 
Engineering Research Council of Canada (NSERC).

\bibliographystyle{h-physrev}
\bibliography{main}

\end{document}